\documentclass[twoside,letterpaper]{soups} 
\usepackage{url}
\usepackage{graphicx, subfigure}
\usepackage{array}
\usepackage{amsmath}
\usepackage{framed}

\begin{document}
%
\conferenceinfo{}{}
\CopyrightYear{2013} 

\title{ChaMAILeon: Exploring the Usability of a Privacy Preserving Email Sharing System.}
%
%
%
%
%

\numberofauthors{1} 
%
\author{
%
%
\alignauthor
$^\dagger$Prateek Dewan, $^\dagger$Niharika Sachdeva, $^*$Mayank Gupta, $^\dagger$Ponnurangam Kumaraguru\\
       \affaddr{$^\dagger$Indraprastha Institute of Information Technology - Delhi, $^*$Delhi Technological University}\\
       \email{$^\dagger$\{prateekd,niharikas,pk\}@iiitd.ac.in, $^*$mayank.gupta@dce.edu}
}

\date{24 July 2013}

\maketitle
\begin{abstract}
While passwords, by definition, are meant to be secret, recent trends have witnessed an increasing number of people sharing their email passwords with friends, colleagues, and significant others. However, leading websites like Google advise their users not to share their passwords with anyone, to avoid security and privacy breaches. To understand users' general password sharing behavior and practices, we conducted an online survey with 209 Indian participants and found that 64.35\% of the participants felt a need to share their email passwords. Further, about 77\% of the participants said that they would want to use a system which could provide them access control features, to maintain their privacy while sharing emails.
To address the privacy concerns of users who need to share emails, we propose ChaMAILeon, a system which enables users to share their email passwords while maintaining their privacy. ChaMAILeon allows users to create multiple passwords for their email account. Each such password corresponds to a different set of access control rules, and gives a different view of the same email account. 
We conducted a controlled experiment with 30 participants to evaluate the usability of the system. Each participant was required to perform 5 tasks. Each task corresponded to different access control rules, which the participant was required to \emph{set}, for a dummy email account. We found that, with a reasonable number of multiple attempts, all 30 participants were able to perform all 5 tasks given to them. 
The system usability score was found out to be 75.42. Moreover, 56.6\% of the participants said that they would like to use ChaMAILeon \emph{frequently}.
\end{abstract}
\vspace{-4pt}

\category{H.1.2}{User / Machine Systems}{Human factors}
\category{H.4.3}{Communications Applications}{Electronic mail}
\vspace{-4pt}
\terms{Human Factors, Design}
\keywords{Email, usable security, password}

\section{Introduction}

A password is a \emph{secret} word or string of characters that is widely used by humans to authenticate themselves against computer systems. Most computer systems rely on passwords to verify an individual's identity, and grant access to the system. It has been well studied that, shared or compromised passwords can lead to serious security implications like identity theft, undue infringement of privacy, security breaches, and misuse of administrative privileges~\cite{Butler:2012,Wilson:2002}. Not surprisingly, the main aspect associated with a password is its secrecy. In the modern world, keeping a password safe, and not sharing it with anyone, features as the top advice in every organization's security policy. Companies like Google, and Yahoo, highly recommend that their users do not share their passwords with anyone.~\footnote{\url{https://accounts.google.com/PasswordHelp}}~\footnote{\url{http://help.yahoo.com/kb/index?page=content&y=PROD_ACCT&locale=en_US&id=SLN3012&impressions=true}}

\begin{figure}[!ht]
\begin{center}
\fbox{\includegraphics[scale=0.205]{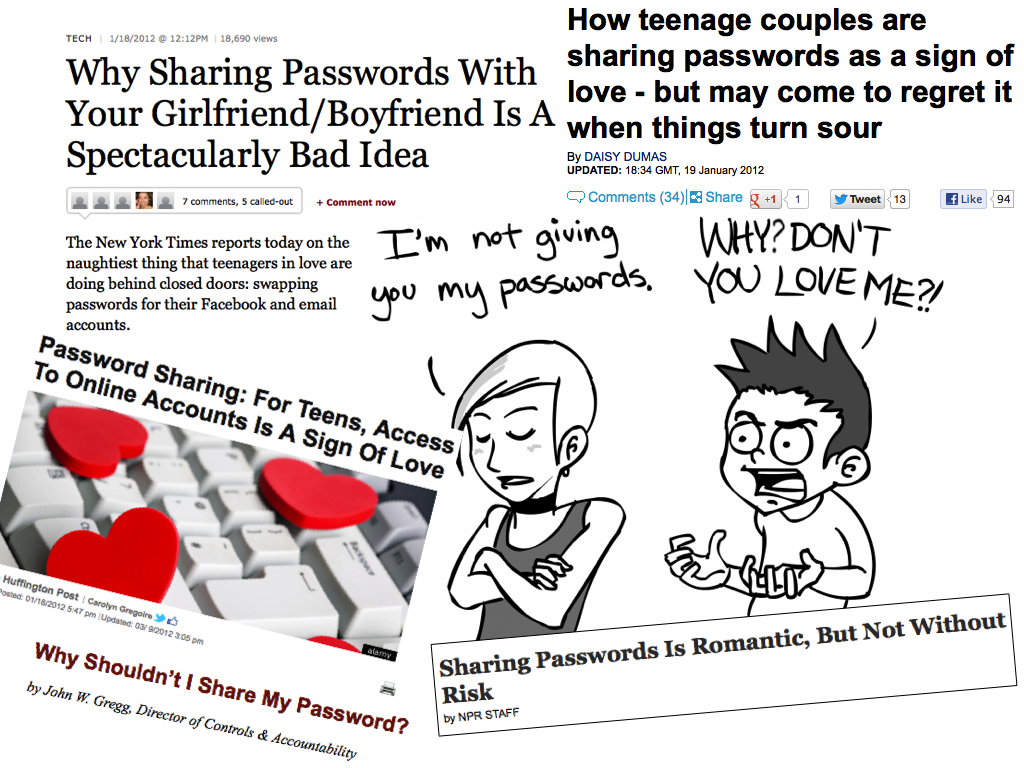}}
\caption{News clips from various international newspapers reporting password sharing among teens as a universal phenomenon.}
\label{fig:front}
\end{center}
\end{figure}

Even though the repercussions of password sharing are well and widely known, it has been observed and studied that, users come across a variety of scenarios in their day-to-day lives, where they share their password with other users. Patrick~\cite{Patrick08monitoringcorporate} used the Enron email corpus to identify instances of password sharing. A study of banking and security in Australia also unveiled the practice of sharing passwords~\cite{Singh:2007:PSI:1240624.1240759}. In order to be more productive, leading executives expressed that they need to share their email accounts with their assistants~\cite{Hyatt:2010}. Sharing passwords amongst teenagers has also become a trend in the recent times (Figure~\ref{fig:front}).

In a recent study, the Pew Internet Research Center found that 1 in 3 teenagers share passwords with their friend, boyfriend or girlfriend~\cite{lenhart2011teens}. 
As reported by The New York Times, some of the implications of sharing passwords include, obsessive monitoring of a significant other's account for signs of infidelity, and using the access for sabotage when a relationship goes sour~\cite{Hill:2012}. These examples clearly indicate that exposing private information while sharing emails is an evident breach of an individual's privacy, which can lead to serious security implications. Users should be able to control which emails they wish to share, and which emails they wish to keep private, when they need to share their email account with their parents, friends, spouse, assistant, or anyone else. Moreover, users should be able to control operations like deletion, and sending of / replying to emails, while they share emails with another user. To the best of our knowledge, there has been some work done to propose methods for sharing emails~\cite{best2010handling},~\cite{gmail-delegation},~\cite{hudecek2005auto},  and~\cite{hwang2012method}, but there exists very little work which caters to the users' privacy needs.

To cater to the users' need for maintaining privacy while sharing~ emails~ and~ email ~passwords, ~we ~introduce ChaMAILeon, a system which allows users to share their email while maintaining their privacy. We present a novel approach of sharing emails, by allowing users to create multiple passwords for a single email address. An email account is thus accessible using more than one password. The user can control which emails and features (reading / sending / deleting) are visible / available if a particular password is used to access her email account. For instance, a supervisor can create 2 passwords, and grant access of her account to her assistant, and her spouse. To maintain privacy, personal emails can be hidden from the assistant, while professional emails can be hidden from the spouse. Our main contributions here are as follows:

\begin{itemize}

\item An online survey with 209 Indian participants to understand users' general password sharing behavior in the Indian context.

\item A web application which allows users to share their emails~ without~~ compromising ~~their ~~privacy. ~ChaMAILeon is a proof of concept for our approach of associating multiple passwords with a single email account, where each password corresponds to a different, restricted view of the email account.

\item A comprehensive evaluation to comprehend the usability, and scope of ChaMAILeon. We conducted a controlled experiment with 30 participants to evaluate the usability of ChaMAILeon.

\item We propose general guidelines for building usable email sharing applications, which provide mechanisms to share emails while maintaining privacy.

\end{itemize}

The rest of the paper is organized as follows. Section 2 discusses the existing work in the area of password practices and email sharing. Section 3 describes the methodology we followed for our online survey and controlled experiment. The system design, and definition of ChaMAILeon are covered in section 4. Section 5 discusses the results of the online survey and controlled experiments. Section 6 summarizes and discusses this work, and explores future scope. The last section contains some general guidelines, which we propose, for building usable and privacy preserving email sharing applications.

\section{Related work}

The practice of sharing passwords, and it's implications have been widely explored in the computer security community. Patrick~\cite{Patrick08monitoringcorporate} used the Enron email corpus to identify instances of password sharing, and found a total of 642 instances of password sharing involving 500 different email addresses. In another work by Alarifi et al., a survey conducted in Saudi Arabia brought out that 35.8\% of 363 respondents shared their access passwords with their family members~\cite{6285845}. A study of banking and security in Australia unveiled the practice of sharing passwords~\cite{Singh:2007:PSI:1240624.1240759}. According to Singh et al., password sharing was seen as a practical way of managing money and a demonstration of trust for married and de facto couples. Sharing Personal Identification Numbers (PINs) is a common practice among remote indigenous communities in Australia. Stanton et al.'s U.S. survey of non-malicious, low technical knowledge behaviors related to password creation and sharing showed that password ``hygiene" was generally poor but varied substantially across different organization types (e.g., military organizations versus telecommunications companies)~\cite{stanton2005analysis}. Their results indicated that 23\% of respondents sometimes reveal their passwords to members of their work groups, 7\% share their passwords with someone in their company but outside their work group, and 4.1\% share their passwords with someone outside their company. Boyd et al.'s work suggests that teens feel safer, and closer to each other by sharing their passwords~\cite{boyd2011social}. 

Apart from the research community, the implications related to password sharing have also caught attention in the news media. Sharing passwords amongst teenagers, has become a trend in the recent times. A recent survey known as `Revenge of the Ex: Love, Relationships \& Technology,' conducted across 10 countries (including USA, Canada, UK, France, Germany, Italy, Mexico, Brazil, Australia and India) revealed that India has the highest percentage of respondents at 43\%, who have shared passwords with past partners.~\footnote{\url{http://articles.timesofindia.indiatimes.com/2013-02-13/pune/37078553_1_passwords-partners-smartphone-users}} However, apart from this study, most of the work in this space focuses on users from the U.S., Europe, and other developed nations. To the best of our knowledge, there exists very little, or no work to capture users' password sharing behavior and practices in the Indian context.

There has also been some work in the industry to cater to the need of sharing passwords, in particular, email passwords. The most recent patent describes a method for generating and managing a user account using a multi-password~\cite{hwang2012method}. This invention specifically targets people in close relationships sharing passwords with each other. Multi-password indicates having multiple active passwords for one account, and therefore users are processed as if the users are logged into different accounts, depending on which password has been inputted. In another patent titled ``Auto-forwarding ~and ~auto-delegating email folder control"~\cite{hudecek2005auto}, inventors describe a software controlled method for auto-forwarding emails to selected recipients, ``enabling a busy executive to speed up forwarding of email messages, delegate their processing and obtain notification of the message processing.'' More patented works include a solution for delegating email messages for human summaries~\cite{best2010handling}. Here, unread emails can be delegated to another email account (belonging to the delegatee), from where, the delegatee can prepare a summary of one or more emails, and send it back to the owner. 

Considering the amount of work done in the research community and industry, real-world solutions for email sharing have been minuscule. Google's mail delegation feature~\cite{gmail-delegation} allows a user to delegate access to her Gmail account to another user / delegatee, so that the delegatee can read, send, and delete messages on behalf of the delegator. When a delegatee sends emails on behalf of the delegator, Google adds an additional ``sent by" field in the email header. This explicitly shows the receiver of these emails, that the email has been sent by the delegatee on behalf of the delegator. However, the delegatee does not get permissions to change her account password or account settings, or chat on her behalf. He also gets full control of her email content. ChaMAILeon, in addition, allows users to control which emails is the delegatee able to read; and to whom is the delegatee able to send emails. This largely increases the application domain of such a system.

Mifrenz~\footnote{\url{http://mifrenz.com/}} is another application, which allows parents to control the email accounts of their children. This application gives parents the ability to let their children safely use email, with the minimum of intervention~\cite{hunt2008mifrenz}. Although, Mifrenz does not directly relate to the problem of password sharing, it overlaps with ChaMAILeon's objective of allowing partial / controlled access to an individual's emails.

Most of the work mentioned above, provides mechanisms to share emails, but does not address any privacy concerns related to email sharing. To the best of our knowledge, ChaMAILeon is the first email application which provides access control in emails. ChaMAILeon specifically targets and caters to users' privacy concerns while sharing emails. Now, we discuss the methodology we followed for conducting the online survey and controlled experiment.

\section{Methodology}

To explore users' password sharing behavior and practices, we conducted an online survey with 209 Indian participants. Based on the insights from the survey, we developed ChaMAILeon. We built a prototype to incorporate features from the users' requirements and concerns, as found in the survey. Even though the data was collected only from India, we believe, the solution we propose is applicable worldwide, for anybody who wants to share their emails without compromising their privacy. Finally, to better explore the system and usage behavior of the participants with the system, we designed and conducted a controlled experiment. 

\subsection{Online Survey}

We first conducted an online survey to study users' general password sharing trends and practices (See Appendix~\ref{append:survey}). The motivation behind conducting this survey was to discover users' intent towards sharing their email passwords, and to see how comfortable they were in sharing passwords. In addition, we also captured if users were willing to use a system which provided them with various access control features on their emails. The online survey was filled by 209 participants from India. The survey was open to everyone, and was publicized on Facebook and through word of mouth. We also sent the survey link to student mailing list of different schools in India. On average, the survey took 15 minutes to complete. Participants were also given the option to participate in a lucky draw by submitting their contact information (email ID or mobile number).

\subsection{Controlled experiment}

\begin{figure}[ht]
\begin{center}
\fbox{\includegraphics[scale=0.215]{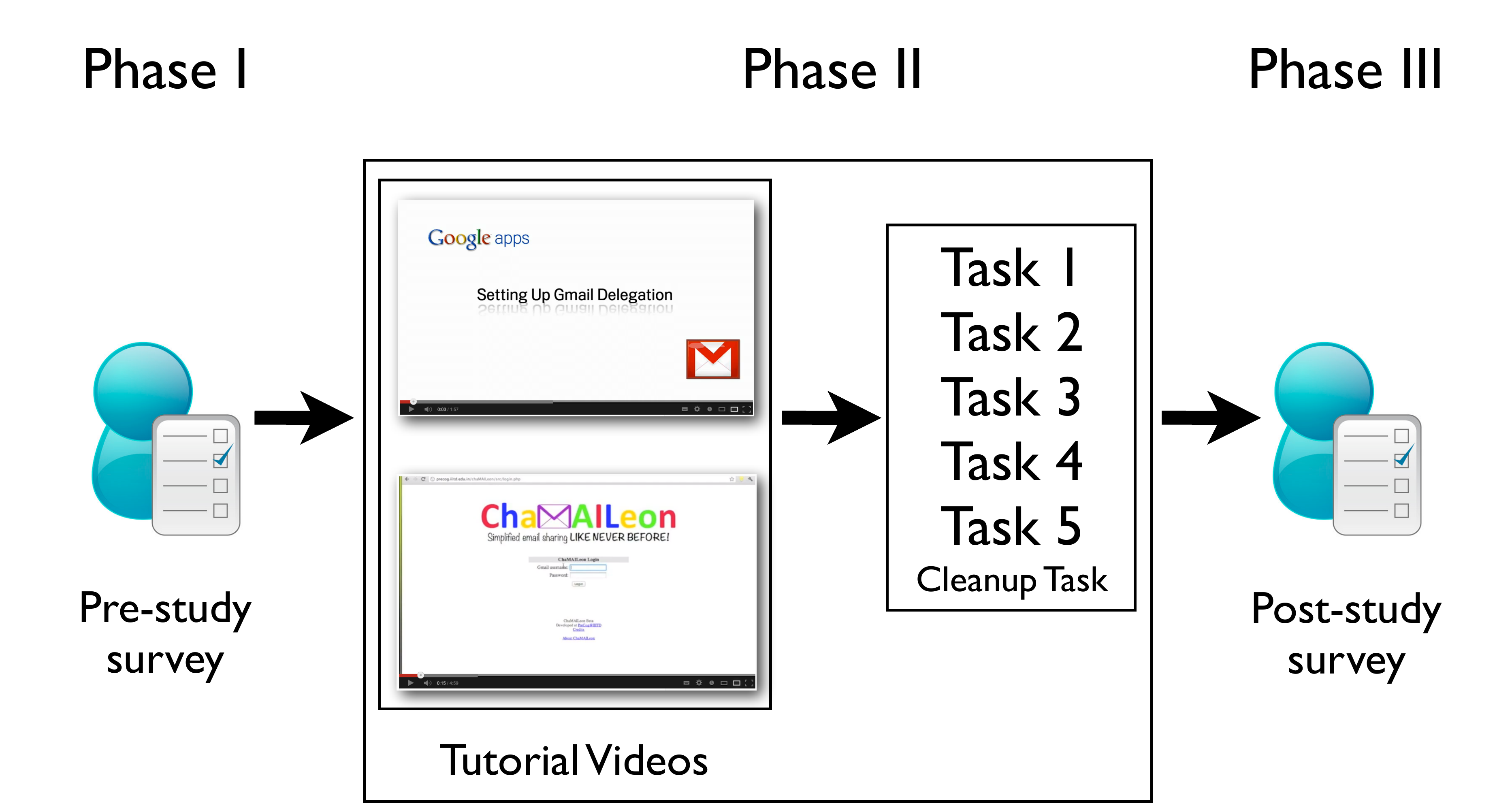}}
\caption{The controlled experiment consisted of 3 phases. Phase I: Pre-study survey, which collected participants' background and demographics, Phase II: Tutorial videos and activity, and Phase III: Post-study survey, which captured participants' experience and feedback for the system.}
\label{fig:studydesign}
\end{center}
\vspace{-15pt}
\end{figure}

\begin{figure}[ht]
\begin{center}
\includegraphics[scale=0.1]{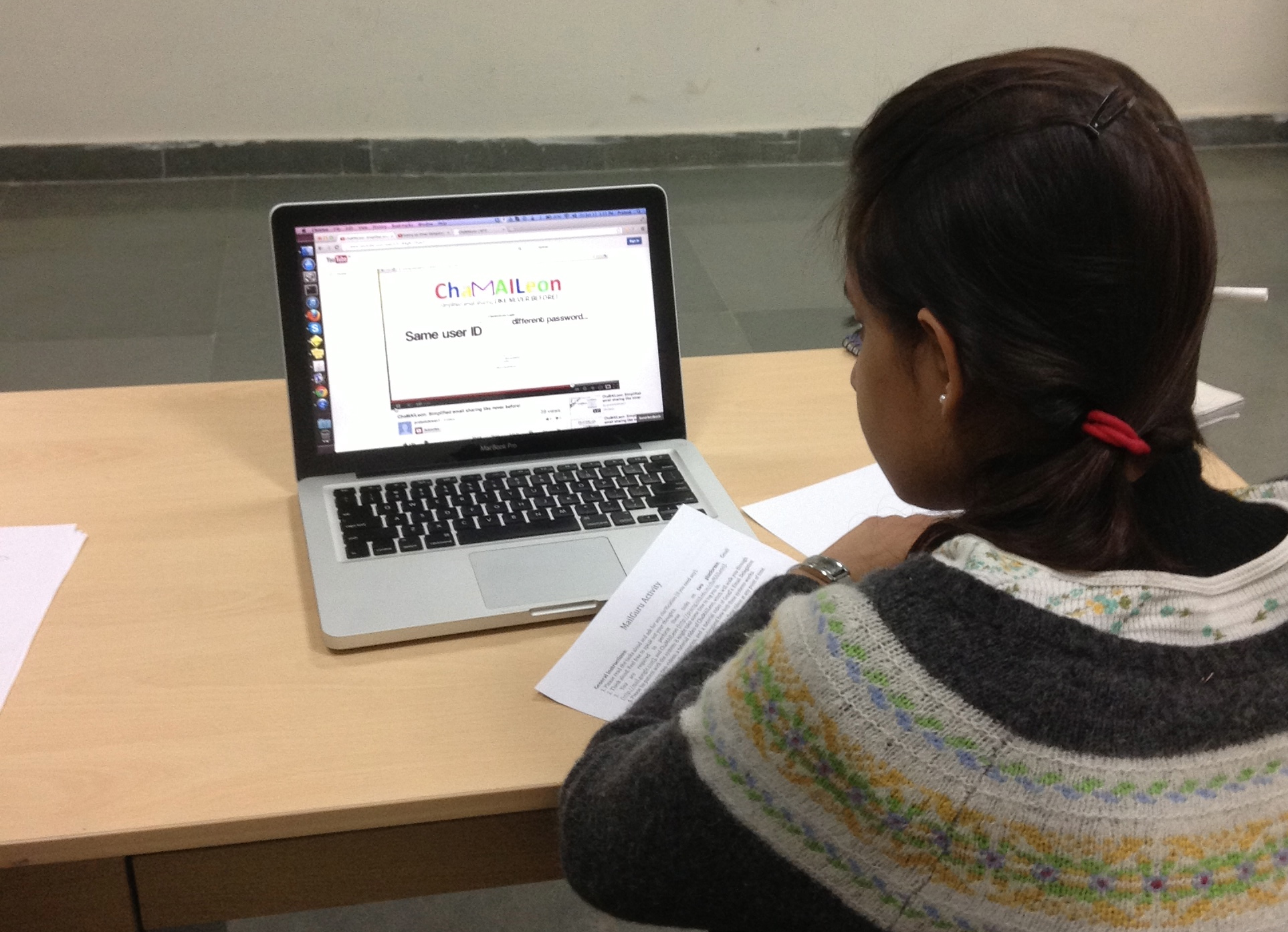}
\caption{Control experiment sessions were conducted inside a lab, one participant at a time.}
\label{fig:userstudy}
\end{center}
\vspace{-18pt}
\end{figure}

We designed a controlled experiment to capture the users' interaction with ChaMAILeon. As depicted in Figure~\ref{fig:studydesign}, the complete experiment consisted of 3 phases; a pre-study questionnaire, an activity involving the user's interaction with ChaMAILeon, and a post-study questionnaire. All the 3 phases were run in the presence of a moderator in a lab at IIIT - Delhi, India. Before starting the activity, all participants were asked to fill the pre-study questionnaire containing 10 questions, along with a consent form, privacy statement, and some general instructions as shown in Appendix~\ref{append:pre}. The activity consisted of a think-aloud session in which participants were asked to play the role of  ``Bob Smith'', the CEO of Bob Construction Company. As part of learning, participants were shown 2 videos, a 5 minute long tutorial video~\footnote{\url{http://www.youtube.com/watch?v=R4gN_GFylzY}} about how ChaMAILeon works and another 2 minute long video titled ``Setting up Gmail Delegation".~\footnote{\url{http://www.youtube.com/watch?v=1I5Xq69E0M8}} We developed counterbalancing schedule to avoid influence of the natural responses of the participants while interacting with the two systems (Gmail and ChaMAILeon); this also helped in reducing the learning effect on the participants. The participants were divided randomly and equally in two groups of 15. Each participant in the group was assigned tasks on Gmail and ChaMAILeon using Latin squares of 2 X 2~\cite{bradley1958complete}. We designed five simple tasks, and asked each participant to perform them on ChaMAILeon, as well as on Google's Gmail to get a comparison between existing techniques and our proposed technique. Figure~\ref{fig:userstudy} depicts the setup of a study session. We provided the participants with a printout that included some general instructions, details about the role, and details and email IDs of people, Bob would be required to interact with, to perform the tasks (as shown in Appendix 2). Participants were free to refer back to both the tutorial videos at any point during the activity.

After the pre-study and the tutorial videos, we asked participants to perform 5 tasks in the activity. These tasks touched upon various aspects of everyday life, where a need for such a system was felt. We gave users the following scenario for performing the tasks: ``Imagine yourself to be Bob Smith, the CEO of Bob Construction Company. You visit your office at 9 am in the morning, and you have 5 tasks to perform in the day, related to your emails.  You may perform these tasks in any order, according to your priority.'' We created a dummy account in the name of Bob, and specified Bob's account credentials with the participants. We further created dummy accounts for various entities which were part of the tasks, and sent some emails from these accounts to Bob's account. The following tasks were required to be performed by the participants:

\begin{itemize}

\item \emph{There are some legal cases currently going on, which you are part of. You (Bob Smith) need to give access to your email to your lawyer Amy. You want Amy to be able to read all your emails, but do not want her to be able to send or delete any emails from your account.} 

\item \emph{You (Bob Smith) want your personal assistant, Penny, to keep a track of emails from a certain set of your foreign collaborators. You thus need to share your email account with her, but wish that Penny reads emails from ONLY collaborator1.mailguru@gmail.com, \newline collaborator2.mailguru@gmail.com, and \newline collaborator3.mailguru@gmail.com.}

\item \emph{You (Bob Smith) would not be able to access your emails for a few hours during the day, and so, you want your colleague Howard to access your emails during these hours. However, you wish that Howard does not get to read emails from your wife (wife.mailguru@gmail.com) ~and ~son (son.mailguru@gmail.com).}

\item \emph{There are some financial emails that you want your accountant Stuart to handle. You (Bob Smith) thus need to give access to your email to Stuart, but you wish that he should be able to see only those emails which contain the word ``accounts'' in the subject line.}

\item \emph{You (Bob Smith) learn that all your appointments for the day have been cancelled, and you realize that you no longer need Howard to access your emails. You want to revoke Howard's access to your account.}

\end {itemize}

To perform a task, participants were required to log into the account as Bob, configure the settings for the sub-user (discussed in the next section), and log out. To verify if the settings made were correct, participants were required to log into Bob's account as the sub-user. This process was repeated for each task, until the participants were satisfied, or gave up. After completion, we asked the participants if they were satisfied with the settings and how different / similar the results were from their expectations. We were not required to go through an IRB-type approval process before conducting the experiment in India. However, authors of this paper have previously been involved in studies with the U.S. Institutional Review Board (IRB)  approvals, and have applied similar practices in this study. Prior to the study, participants were shown a consent form, which they had to read and accept, if they were comfortable with it. The form stated that an audio recording and a screencast would be taken during the activity, and that the collected data would be anonymized and used only for the purpose of this research. Furthermore, participants were informed that they could withdraw from the study at any point and request the deletion of the audio recording and screencast.

The controlled experiment was followed by a post-study feedback questionnaire with 11 questions. This comprised of 10 standard SUS questions~\cite{brooke1996sus} and 1 optional open feedback field.

\subsection{Participants}

~\textbf{Online Survey:} Participants were recruited through Facebook posts, student mailing lists, and word of mouth. In all, 209 participants from various ethnic backgrounds participated in the survey. Table~\ref{table:soups2013:demographics} discusses the demographics of the participants from the online survey. The participants used various email services. 97.60\% were Gmail users, 49.76\% were using Yahoo Mail, 26.31\% were Hotmail users and 9.56\% were Rediffmail users. 
The male population comprised of 66.5\% of the participants, whereas 33.5\% were females. Authors of this paper recently conducted a large-scale privacy study across all states in India, with over 10,000 participants, and reflected very similar male : female ratio; which helps us to safely conclude that our sample is representative of a bigger population~\cite{kumaraguru:privacy-in-india:-attitud:2012:yuqfj}.

\begin{table}[!h]
\centering
\caption{Demographics of online survey participants; (N $=$ 209). }\label{table:soups2013:demographics}
\begin{tabular}{|p{5cm}|p{2.55cm}|} 
\hline
\bf{Gender} &  (In \%)\\
\hline
Female &33.50 \\
\hline
Male &66.50\\
\hline
\hline
\bf{Age} (in years) &  (In \%) \\
\hline
 13 to 18 & 18.66\\
 \hline
 19 to 25 &73.68\\
 \hline
 26 to 32 &5.27\\
\hline
 33 to 40  &0.48 \\
\hline
 41 to 50 & 1.43\\
 \hline
Above 50 & 0.48 \\
\hline
\hline
\bf{Education Completed} &  (In \%)\\
\hline
 High school& 45.94 \\
\hline
Under graduation& 35.89\\
\hline
Post graduation & 15.79\\
\hline
 Ph. D. & 1.43\\
 \hline
 Other & 0.95\\
 \hline
 \hline
\bf{Occupation / Profession} &  (In \%)\\
\hline
 Computer Science&44.98 \\
\hline
Engineering&37.80\\
\hline
Business &2.87\\
 \hline
Finance &2.39\\
\hline
 Other &11.96\\
 \hline
\end {tabular}
\label{table:soups2013:demographics}
\vspace{-12pt}
\end{table}

~\textbf{Controlled study:} We recruited participants through advertisements on mailing lists, and word of mouth.  Each participant was offered a monetary reward of 100 Indian Rupees (approx. USD 2) for participating in the study. Most of the participants in the controlled experiment were graduate and undergraduate students and were heavily using emails as a medium of communication. Participants in the control experiment had a decent computer science and technology exposure. 
Majority of the participants belonged to 18-25 age group as this group constitutes the majority of users who would need an application like ChaMAILeon. Table~\ref{table:soups2013controll:demographics} summarizes the demographics of the participants from the control experiment.

\vspace{-0.1in}
\begin{table}[!h]
\centering
\caption{Demographics of controlled experiment participants; (N $=$ 30). }\label{table:soups2013controll:demographics}
\begin{tabular}{|p{5cm}|p{2.55cm}|} 
\hline
\bf{Gender} &  (In \%)\\
\hline
Female &53.33\\
\hline
Male &46.67\\
\hline
\hline
\bf{Age} (in years) &  (In \%) \\
\hline
 18 to 25 & 96.67\\
\hline
 26 to 32 & 3.33  \\
\hline
\hline
\bf{Education Completed} &  (In \%)\\
\hline
School / High school&43.33 \\
\hline
Undergraduation& 36.67\\
\hline
Post Graduation &20.00\\
\hline
 Doctorate&0.00\\
\hline
\end {tabular}

\label{table:soups2013controll:demographics}
\vspace{-10pt}
\end{table}

\section{System design}

\begin{figure}[ht]
\begin{center}
\fbox{\includegraphics[scale=0.27]{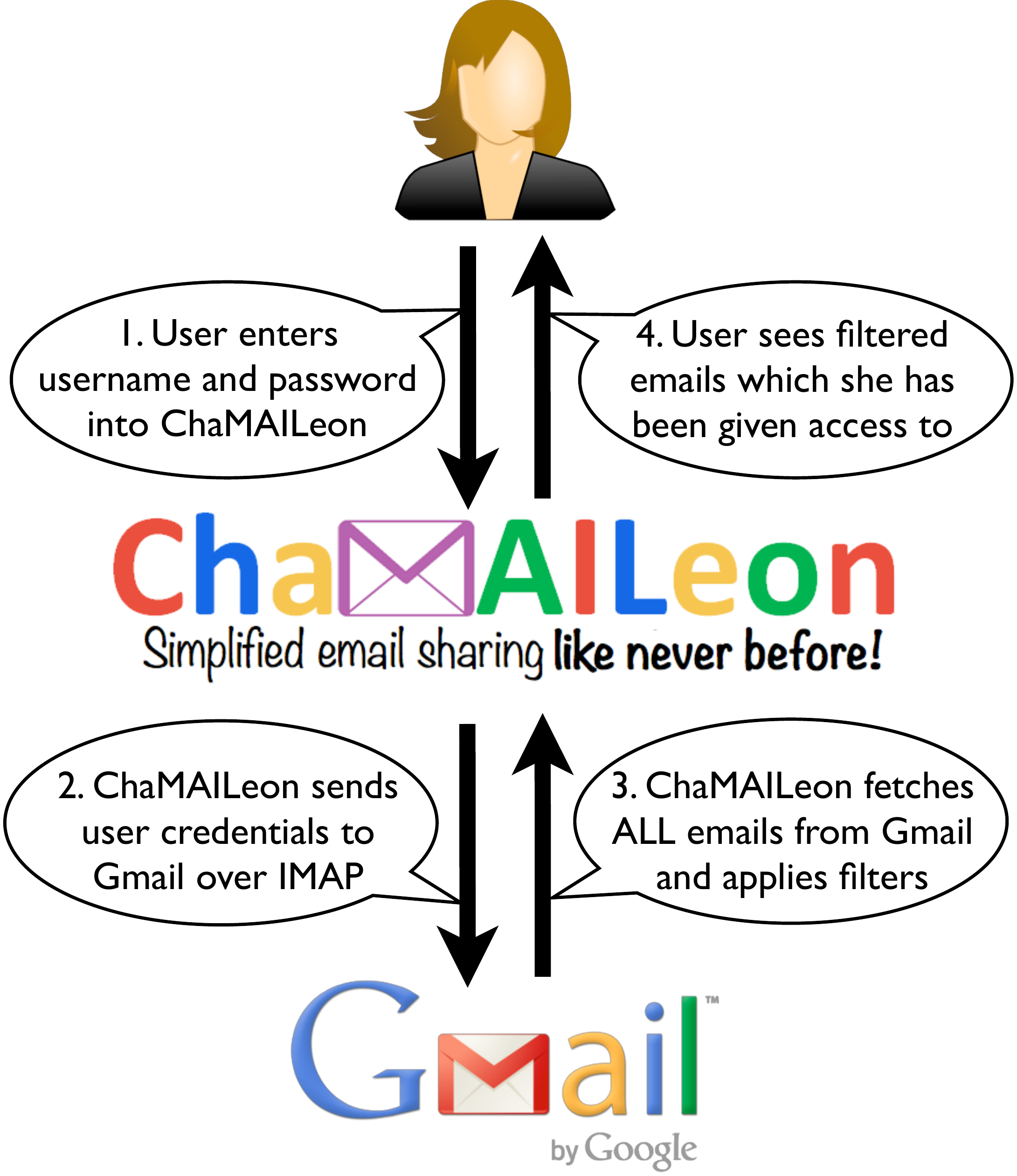}}
\caption{Flow diagram depicting how ChaMAILeon works as a middleware between a user and Gmail's mail server.}
\label{fig:ch_arch_3}
\end{center}
\vspace{-20pt}
\end{figure}

In response to the results of the online survey (discussed in section 5.1), we developed ChaMAILeon, an email application which provides users with the features described in the survey scenarios. ChaMAILeon is a modified version of SquirrelMail~\footnote{\url{http://squirrelmail.org}}, which is an open source email package built using PHP. SquirrelMail was altered to support a ``Configure Account" option to enable users to set access control over their email accounts. ChaMAILeon can be configured to work with any of the existing email services, like Gmail, Yahoo! Mail, Hotmail etc., or an organization's own proprietary mail service. In our case, we configured ChaMAILeon to work with Gmail. Since ChaMAILeon makes use of Simple Mail Transfer Protocol (SMTP) and Internet Message Access Protocol (IMAP), it is fairly simple to configure the system to work with any email service which makes use of these protocols. Apparently, configuring ChaMAILeon is very similar to configuring email clients like Microsoft Outlook, Mozilla Thunderbird, Apple Mail etc., and supports all email services which these clients can support. Figure~\ref{fig:ch_arch_3} represents how ChaMAILeon communicates between a user and Gmail.

\begin{figure}[!t]
\begin{center}
\fbox{\includegraphics[scale=0.32]{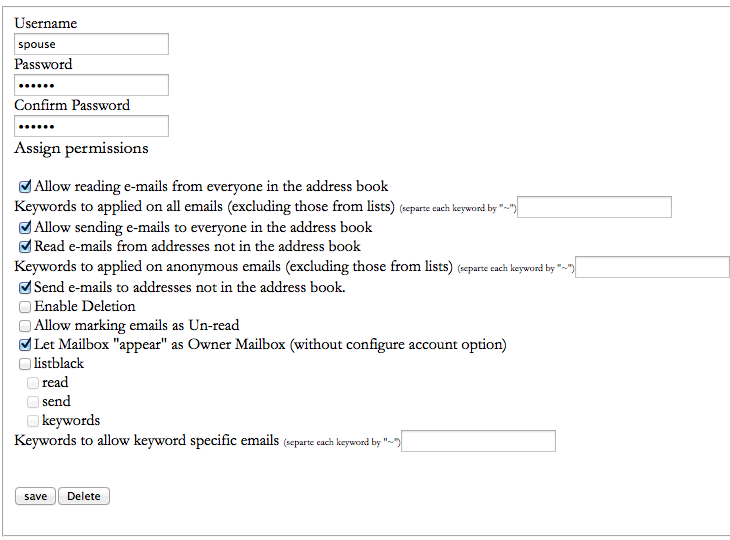}}
\caption{ChaMAILeon's ``Create sub-user'' page. It can be used by the owner of an email account to create another password and allow partial access of her account to someone.}
\label{subuser}
\end{center}
\vspace{-20pt}
\end{figure}

Technically, ChaMAILeon adds two new features to the existing web mail technology, which alters the way the conventional email works. First, it allows the owner of the email account to create ``sub-users" for their email account. A ``sub-user" essentially corresponds to a new password \emph{$P_i$}, which can be used to~ access~ this~ email ~account (Figure~\ref{subuser}). ~ChaMAILeon thus allows to log into one email account using one email ID \emph{E}, and multiple passwords \emph{$P_1$, $P_2$, $P_3$...} Each password (or sub-user) \emph{$P_i$} can be configured by the ``owner" of the email account, to provide different views \emph{$V_i$} of the email account and grant different permissions. This can be viewed similar to roles in Role Based Access Control in databases~\cite{ferraiolo1995role,sandhu1996role,zhang2003pbdm}. In contrast to databases, however, ChaMAILeon identified the access controls using passwords instead of user-names. Each ``sub-user'' can be thought of as a role, and gets a different level of access to an email account. The various rules that can be configured are:



\begin{enumerate}

\item Allow reading emails from IDs present in Contacts (optional keyword filter for these emails)
\vspace{-7pt}
\item Allow sending / replying emails to IDs present in Contacts 
\vspace{-7pt}

\item Allow reading emails from IDs not present in Contacts (optional keyword filter for these emails)
\vspace{-7pt}
\item Allow sending / replying emails to IDs not present in Contacts 
\vspace{-7pt}

\item Enable deletion of emails
\vspace{-7pt}
\item Allow marking emails as ``unread''
\vspace{-7pt}
\item Let mailbox ``appear'' as owner's mailbox
\vspace{-7pt}
\item List specific permissions (read, send, keyword filtering)

\end{enumerate}


Secondly, ChaMAILeon provides the owner of the account, with an option to create ``lists''. Each list is simply a collection of email addresses clubbed together under one name; the name of the list. The advantages of creating such lists are realized through configuring the sub-users. Each list can act as a blacklist or a whitelist for a sub-user, i.e. a sub-user can be allowed / restricted to ``send to" and ``read emails" from a ``list" of email addresses specified in the list.

A combination of sub-users and lists form a powerful and robust mechanism to generate and implement fine-grained access control rules on emails. For instance, consider a scenario where Alice, a company supervisor, wants to share her email with her personal assistant Bob. Alice wants that Bob should log into her email account, and read / respond to only those emails which contain the word ``office" in their subject line. However, Alice wants that Bob should not be able to see any of her personal emails sent to her from her husband (husband@somemail.com), and her daughter (daughter@anemail.com). Also, Alice does not want that the receivers of emails sent by Bob on her behalf, should come to know that the emails have not been sent by Alice herself, but by Bob. Alice does not want Bob to be able to delete any of her emails either.

To the best of our knowledge, none of the existing web mail services provide mechanisms to do this. However, this can be easily set up using ChaMAILeon. Alice can simply add her husband's and daughter's email addresses in a list, and create a sub-user for Bob, which does not have access to this list. Alice can also specify the keyword ``office" in the keyword filter that may be applied to all other emails. Other options like allowing to send emails, and disabling deletion can be applied by simply checking a check-box.

ChaMAILeon was set up on an IBM server with Intel Xeon processor, and 32 Gigabytes of RAM.~\footnote{Live at \url{http://precog.iiitd.edu.in/chaMAILeon}} The system uses a MySQL database to store users' Gmail passwords in an encrypted form, their imported contacts, sub-user details, lists, and usage logs. Access control rules on emails are enforced by parsing the entire set of the user's emails fetched from the Gmail server. Restricted emails are identified based on the sender's address and keywords, as defined by the user. Once these emails are identified, they are removed from the set of emails which are presented to the user on the mailbox interface page. Features like restriction on deletion of emails, composing new emails, replying to emails, etc. are implemented by removing the ``buttons" responsible for these actions. Implementing all of the aforementioned features only requires basic knowledge of PHP programming language, and MySQL.

\subsection{System definition} \label{sysdef}

\begin{figure}[ht!]
\begin{center}
\fbox{\includegraphics[scale=0.27]{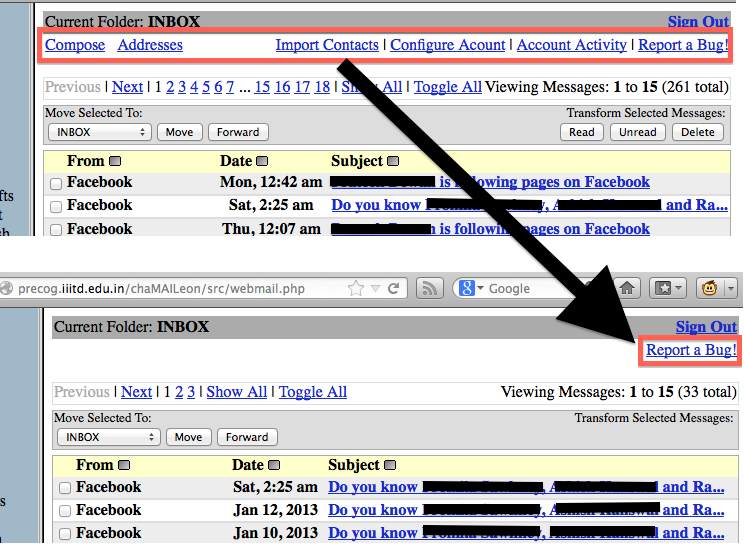}}
    \caption{%
        The sub-user does not get to see all features, which are otherwise present in the owner's inbox. The highlighted area in the top image is different from that in the bottom image.
     }%
   \label{fig:diff}
\end{center}
\vspace{-17pt}
\end{figure}

A user logging into ChaMAILeon can have 2 different roles, viz. \emph{the owner}, and \emph{the sub-user}. When a user logs in as the owner, she gets full control of the email account, and the user interface includes features like \emph{Import contacts}, \emph{Configure account}, \emph{See account activity}, and \emph{Report a bug}. We call this interface as the \emph{``owner's interface"}. However, except for \emph{Report a bug}, all other features are not present in the user interface when a user logs in as a sub-user (Figure~\ref{fig:diff}). In the \emph{add / edit sub-user} page (Figure~\ref{subuser}), we also give an option to the owner, to \emph{``let mailbox look like the owner's mailbox"}. Enabling this option adds dummy buttons for ``Compose", and ``Addresses" to the sub-user's interface. These buttons are non-functional. Clicking either of them points back to the user's inbox. This is done to avoid a sub-user from getting a non-familiar feeling by not finding these essential features in a mailbox user interface.

We now formally define the parameters associated with each instance of a sub-user $S$ as a 10-tuple vector, where each vector represents a specific permission.

\vspace{5pt}
\begin{centering}

$S$ = param<$R_c$, $R_{c}^k$, $S_c$, $R_{nc}$, $R_{nc}^k$, $S_{nc}$, $D$, $U$, $Z$, $L$>

\end{centering}
\vspace{5pt}

\noindent where:
\vspace{-6pt}
\begin{table}[h]
    \begin{tabular}{p{11pt}p{0.5pt}p{7cm}}
    
        $R_c$          & = &Read emails from contacts                                                                      \\ 
        $R_{c}^{kc}$   & = &Read email from contacts, containing \emph{keywords} $kc$ in the subject line            \\ 
        $S_c$          & = &Send / reply emails to contacts                                                                        \\ 
        $R_{nc}$       & = &Read emails from non-contacts                                                                  \\ 
        $R_{nc}^{knc}$ & = &Read email from non-contacts, containing \emph{keywords} $knc$ in the subject line\\ 
        $S_{nc}$       & = &Send / reply emails to non-contacts                                                                    \\ 
        $D$            & = &Delete emails                                                                                  \\ 
        $U$            & = &Mark emails as unread                                                                          \\ 
        $Z$            & = &Spoofing (Making interface ``look" like the owner's interface)                                 \\ 
        $L$            & = &\emph{name1}<$R$, $S$, $K$>, \emph{name2}<$R$, $S$, $K$>, ... 

    \end{tabular}
\vspace{-20pt}
\end{table}

All vectors except $L$ are binary vectors and take only two possible values, i.e. \emph{1 - enabled}, or \emph{0 - disabled}. List permissions $L$ are an array of 3-tuple vectors, where $R$, and $S$ respectively represent \emph{read-from} permissions, \emph{send / reply to} permissions, and $K$ is a \emph{set} of keywords. $R_i$ and $S_i$ are binary vectors.

\section{Results}
In this section, we will discuss the results of the online survey and controlled experiment.
\subsection{Online survey}

Our online survey was taken by 209 Indian participants. To the best of our knowledge, this is the first work on studying password sharing perceptions and practices in India.

\subsubsection{Current user behavior and expectations}
Participants showed contradictory behavior to what was recommended by popular email services (Gmail, and Yahoo). We asked participants, if they share or feel the need for sharing passwords of their email accounts, and found that 64.35\% of the participants felt a need for sharing email passwords. We observed that 97.60\% of all participants were using Gmail, and 49.76\% were using Yahoo Mail. To support our findings, we visited the security policies of these two email services, and found that both of them highly recommended their users not to share passwords.
%

We asked participants how comfortable they were in sharing their passwords with one or more people, and found that majority of the participants (74.64\%) were not comfortable sharing passwords. Further, 38.75\% were \emph{very uncomfortable} with sharing passwords. This indicated a need for systems which allow privacy preserving email password sharing facilities to the users.

\subsubsection{Fears, worries and actions}
Password sharing leads to various apprehensions and worries for the security of the email account in users. 
We wanted to investigate the post effects of password sharing in the context of Indian users. To understand this, we asked if participants thought password sharing might lead to unwanted consequences. We found that 88.51\% of the participants thought password sharing may lead to unwanted consequences. In fact, 11.48\% of these participants mentioned that they had to face unwanted consequences as a result of sharing their password with someone.

To understand the user apprehensions, we asked participants, if they would consider changing their password after sharing it. We found that 96.65\% of the participants felt the need to change their passwords after sharing. Among these, 45.93\% mentioned that they would definitely change the password immediately, and 35.40\% felt that they would probably change it whenever convenient. 15.31\% of the participants mentioned that they may consider changing it some time. These results indicated user worries and uneasiness when they share their passwords, and reasonably signify the need for systems which could provide users with easy and fearless email sharing.

\subsubsection{Need for a system}
To better understand the exact system utilities, features and requirements, we gave participants a scenario, in which a system allows them to control the accessibility of their emails, settings, and features, when they share their email password. We then gave participants 3 scenarios as mentioned in Appendix~\ref{append:survey}. 
 
\begin{figure}[ht]
\vspace{-10pt}
\begin{center}
\includegraphics[scale=0.45]{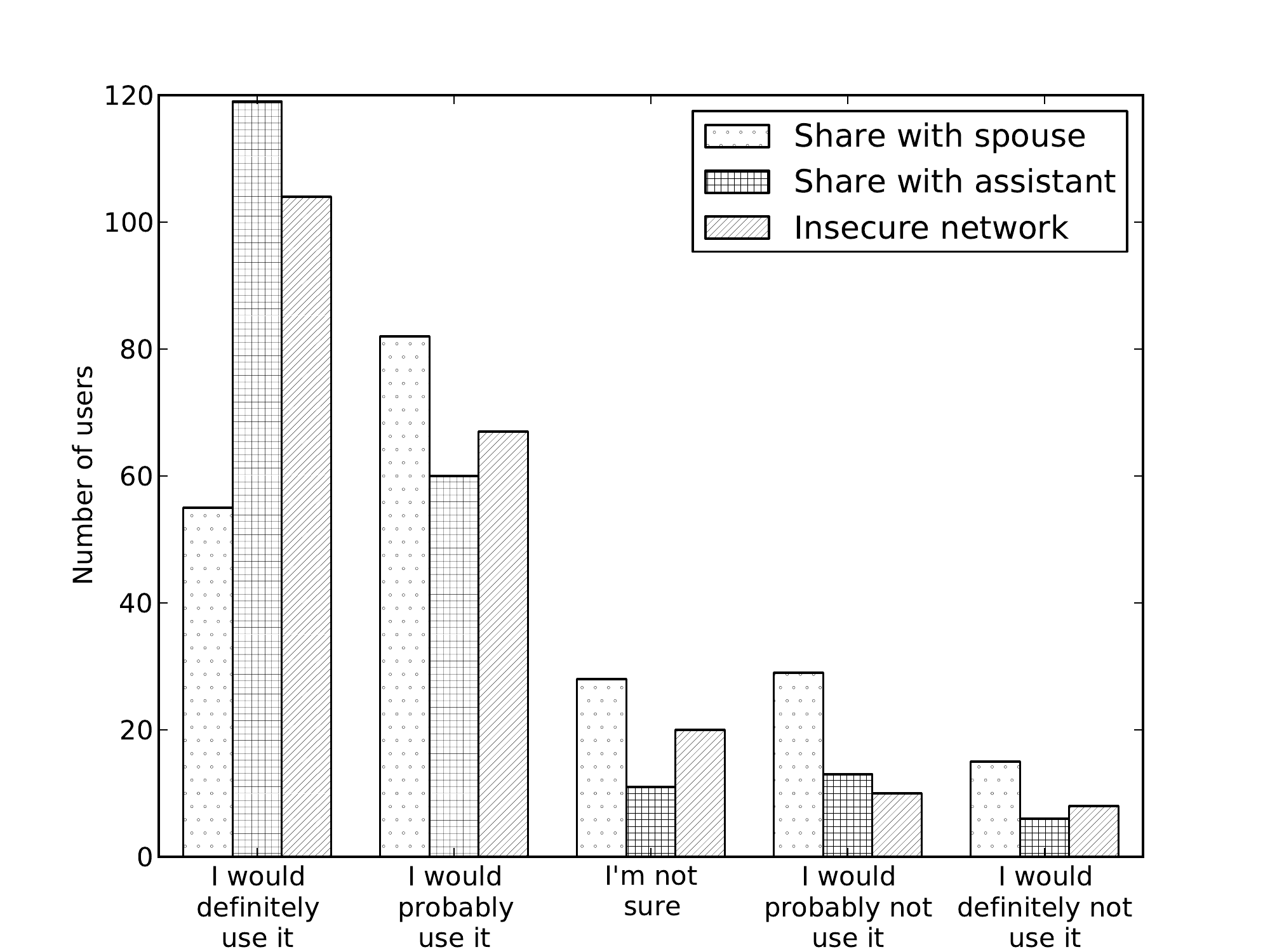}
\caption{The need for a system like ChaMAILeon was felt the most in professional environments. A big fraction of participants (85.64\%) said that they would want to use such a system while sharing their emails with their assistant.}
\label{nfs}
\end{center}
\vspace{-20pt}
\end{figure}

\textbf{Share emails with spouse:} We asked participants to consider a situation, where they wanted to share their email with their spouse, but did not want him / her to see emails from a particular friend, or did not want him / her to be able to reply to any emails. We then asked them if they were willing to use a system which could provide them with this functionality. We found that 65.55\% of the participants were willing to use such a system. Moreover, 26.31\% opted for \emph{definitely} using such a system (See Figure~\ref{nfs}).
 
\textbf{Share emails with assistant:} In the next scenario, we touched upon the need of such a system in a professional environment. We asked participants to consider a scenario where they were an official at a high post in a company, and received many emails every day. They wished to allow their assistant to be able to log into their account and reply to emails on their behalf. However, they would want the assistant to be able to see emails sent from only specific people, and not all of them. We found that participants felt the need for such a system more in a professional environment, as compared to the previous scenario (See Figure~\ref{nfs}), with 85.64\% agreeing to use of such a system in the current scenario.
 
\textbf{Accessing emails in an insecure environment:} In the last scenario, participants were enquired about the need for such a system in not so secure environments or untrusted networks. We asked, if participants were in a cyber cafe where they were not sure if the network was secure, and they needed to check their email. In such a setting, users would be afraid that their password and email content might get compromised over the insecure network. Under such a scenario, they were given an option to log into their account with a different password, such that, logging in with this password would give them very limited / restricted access to their account. In addition, their important emails would be filtered, and would not be transmitted to them over this insecure / untrusted network. 81.82\% of the participants mentioned that they would use such a system (See Figure~\ref{nfs}).

\subsection{Controlled experiment}
We conducted a controlled experiment with 30 participants. Including the pre-study questionnaire, screening of two videos (5 minutes + 2 minutes), 5 tasks, and a post-study questionnaire, the experiment took about 45 minutes for each participant to complete. All participants were asked to perform all 5 tasks on both Gmail, and ChaMAILeon. Apparently, 20 out of 30 participants had never heard of Gmail's delegation feature before. Further, only 12 out of the 30 participants attempted at least one task on Gmail. The other 18 participants stated that none of the 5 tasks (mentioned in section 3.2) could be performed on Gmail. Also, none of the 12 participants who attempted one or more tasks on Gmail, could achieve success in any of the tasks they attempted. On the contrary, all 30 participants were able to complete all 5 tasks on ChaMAILeon, with an average of 1.33 attempts per task per user.

\subsubsection{Password sharing practices and trust}

\begin{figure}
\begin{center}
\includegraphics[scale=0.5]{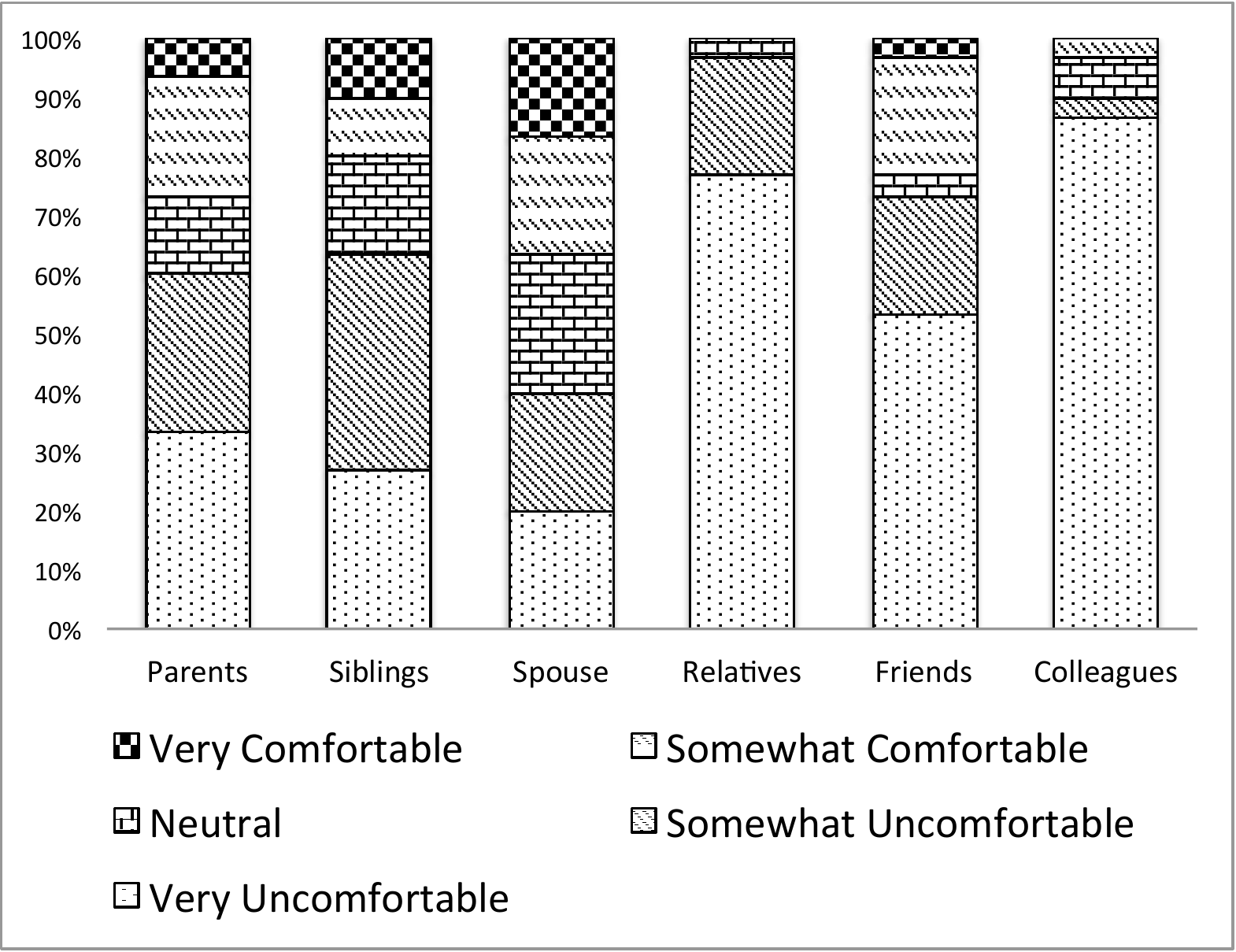}
\caption{Comfort levels of participants in sharing email passwords with various entities. Users were found to be least comfortable in sharing email passwords with their colleagues.}
\label{fig:pre_relations}
\end{center}
\vspace{-22pt}
\end{figure}

From the pre-study questionnaire, we found that 40\% of the participants in the control study shared their passwords with others, and 80\% maintained different email accounts for personal and professional needs. To analyze the level of comfort and trust participants experienced while sharing their passwords with different people, we asked participants how comfortable they felt in sharing passwords with parents, siblings, spouse, relatives, friends and colleagues. We found that participants were most uncomfortable sharing passwords with colleagues (86.67\%). This supports our results from the online survey which suggested that a majority of users wanted to use systems like ChaMAILeon to share their emails with their assistants, or in professional environments. We also found that participants felt very uncomfortable sharing passwords with friends (53.34\%) and parents (33.34\%) as shown in the Figure~\ref{fig:pre_relations}. These results indicate towards a need for a system which enables role based password sharing and access to emails.

\subsubsection{Participants' Experience and Interface Feedback}

After the experiment, we asked participants a set of 10 questions to evaluate the usability of ChaMAILeon~\cite{brooke1996sus}. The System Usability Score (SUS) came out to be 75.42. It can be interpreted as a grade of a B-. A score of above 80.3 is required to get an A. This is also the point where users are more likely to be recommending the product to a friend.~\footnote{\url{http://www.measuringusability.com/sus.php}} We conclude with this score, that ChaMAILeon is fairly usable. The evaluation process and user feedback helped us to identify the various issues with the interface, and possible scope of improvement.

We also asked participants an open ended question about their comments and feedback for the system. Despite pointing out some interface issues, 83.3\% of the participants seemed to be happy with their experience with the system, and stated that they would like to use such a system regularly. Participants also stated that such a system was really needed. One participant quoted, ``Nice and very well needed system since almost everyone has a Gmail account these days and need to share their password at times." Participants seemed to like the idea very much, and said, ``Apart from minor interface issues, the system is user friendly and the idea is amazing, and very practical, as well as easy to get used to. Thumbs up!" ChaMAILeon's use in the professional environment was also pointed out by one of the participants when they said, ``The system is an excellent initiative which can prove to be very useful for corporate organizations." Some overwhelming responses included a participant saying, ``After using this system I am thinking why this system is not so popularized. It is very easy to use, all functionalities are well defined, so why should I use any other mail system instead of ChaMAILeon? After some time I can say surely that it will be the 1st popular mail system. I enjoyed by playing with this system." A quote from another participant was, ``Great feature of hiding your personal mails and still getting your work done!!! Would love to use this system."

There were 5 participants (16.6\%) who found the interface non-intuitive and hard to use, and stated that, ``User interface could be better." Another such participant said, ``Very useful, but higher learning curve than expected / desired. Can be easy to use with better UI and more clear options. Definitely an edge over Google's delegate function." One participant was more specific with the issues she faced with the interface, and stated, ``Checkboxes and a delete button around the place where the users and lists are given, would be of great help." We will now discuss the most common experiences and interface feedback by the participants in detail.

\paragraph{Common experiences} We found that, on multiple occasions, participants forgot the passwords of the sub-user accounts they created. This phenomenon of users forgetting passwords is well established where users have multiple accounts, and are required to remember multiple passwords~\cite{notoatmodjo2009passwords}. As a solution for this problem, in ChaMAILeon, a password generation and management system can be provided to the user, which can generate and store secure passwords for various sub-users. These passwords would be visible only to the owner, and she would not have to remember all passwords.
Few participants were found to be using the same passwords for two different sub-user accounts. This lead to two different permission tuples being mapped to a single password, which eventually confused the system, and resulted in a failed attempt. Participants suggested that the system should prompt if a user tries to keep the passwords for two different sub-user accounts. Participants were also found to use interesting names for the lists and sub-users they created. While most participants named the sub-user accounts as ``Amy", ``Penny", ``Howard", and ``Stuart", some participants used names according to tasks, (like ``Task 1", ``Task 2" etc.), and roles (like ``Accountant", ``Lawyer" etc.). Similarly, participants used various conventions for naming lists. Most participants named the two lists as ``collaborators", and ``family", while others named lists according to the sub-users. For instance, a few participants named the list of collaborators as ``Penny", since Penny was supposed to be given access to this list. Other names included ``foreign\_collab", ``personal", ``secret", ``relatives" etc. Observing the arbitrary conventions, we infer that automatic email labeling and classification techniques can be used to generate lists automatically~\cite{bekkerman2004automatic,crawford2001automatic,provost1999naive}.
Almost all participants referred to the tutorial video during the tasks. Since this was the first interaction of the participants with ChaMAILeon, we expected participants to seek help. It was interesting to find that using only the tutorial video as helping material, all participants were able to successfully complete all tasks.

\begin{figure}[!h]
\vspace{-5pt}
\begin{center}
\fbox{\includegraphics[scale=0.25]{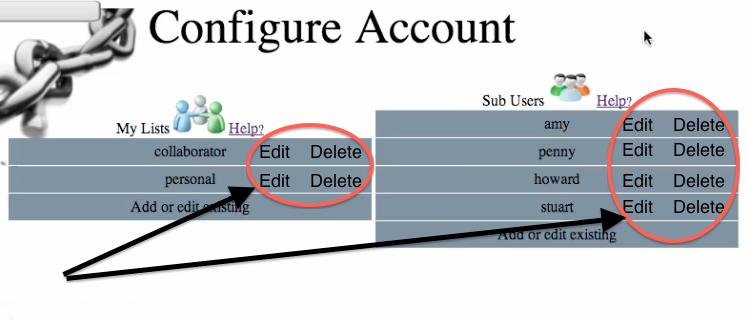}}
\caption{A snapshot of the ``Configure Account" page after completion of Task 4. Participants proposed to have ``Edit" and ``Delete" buttons for lists and sub-users at the areas marked and pointed in this image.}
\label{fig:editdeletebuttons}
\end{center}
\vspace{-15pt}
\end{figure}

\paragraph{Interface feedback} Most participants found the interface reasonably usable, but suggested various changes to make the system more intuitive. For Task 5, which required the deletion of a sub-user account in order to revoke access, most participants found it surprisingly hard to locate the ``Delete" button on the sub-user page. This button was placed right next to the ``Save" button on the sub-user page, and all participants used the ``Save" button multiple times while performing the tasks. Surprisingly, it seems like most participants never noticed the ``Delete" button placed right next to the ``Save" button. As a result, participants consumed a lot of time trying to find how to delete a sub-user, which lead to an increase in the average completion time for this task. The most common suggestion to fix this issue was to have an ``Edit", and a ``Delete" button next to each sub-user and list visible on the ``Configure Account" page (as shown in Figure~\ref{fig:editdeletebuttons}), instead of having to go to each sub-user and list, and delete it.

Setting permissions for lists required participants to check / uncheck the checkboxes corresponding to ``Read", ``Send", and ``Keywords" permissions under each list created by the participant. All checkboxes were unchecked by default. During Task 3, participants found it clumsy to check all permissions corresponding to each list separately, and stated that it would be a tedious process if the number of lists were large. Participants suggested to have all list permissions checked by default, and preferred to uncheck the lists which needed to be blacklisted. A better suggestion was to have separate ``pools" for black lists and white lists, with all lists in the whitelist pool by default. Participants then proposed a ``drag-and-drop" approach to ``move" lists from the pool of whitelists to the pool of blacklists, as required.

Further, in order to ensure that the settings made for each task were satisfactory; participants had to log out after making the settings, and log back in as the sub-user that they wanted to verify the settings for. Participants complained that this process was tedious, and suggested to have a ``preview" pane within the sub-user settings page, which would contain the dynamic view of each sub-user's view of the email inbox.

\subsubsection{Task Completion Time}

On an average, a participant took 17 minutes and 53 seconds to perform all 5 tasks (min. 10:17, max. 37:17). However, the complete experiment session (excluding pre-study and post-study), from first sign in to the last sign out, took 32 minutes and 35 seconds on average, for each participant. The difference between the two time values can be attributed to the time spent on the cleanup task, verification of each task, attempts made on Gmail, and reading the tasks from the handouts. The time taken for each task was calculated on the basis of the session duration i.e. time between a participant logging in as Bob, and logging out. Session durations for all such sessions, which were required to successfully complete one task, were aggregated to get the final duration of time that was taken for the completion of a task. We performed a two-tailed Student's T-Test, and did not find any statistically significant difference in the total time taken between the group of participants who viewed the Gmail delegation video first, and the group of participants who viewed the ChaMAILeon tutorial video first (2 sample t test, p-value $>$ 0.1).

\begin{figure*}[!ht]
\begin{center}
\includegraphics[scale=0.367]{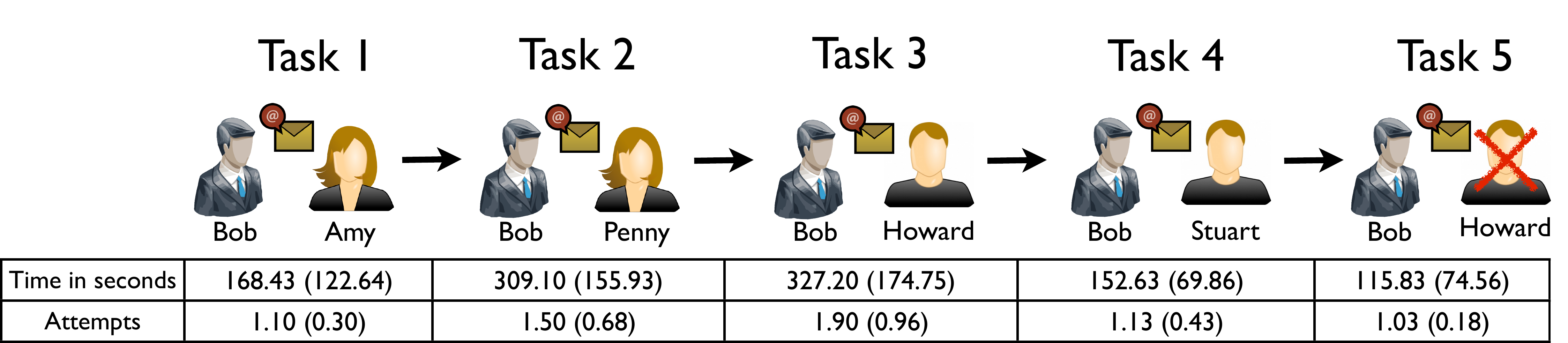}
\vspace{-8pt}
\caption{Average completion time and average number of attempts per task. The values within braces represent the standard deviation. (i) Amy (lawyer) is allowed to read all emails; (ii) Penny (assistant) is allowed to read emails from collaborators only; (iii) Howard (colleague) is not allowed to read personal emails; (iv) Stuart (accountant) is allowed to read emails related to accounts only; and (v) Bob revokes Howard's access.}
\label{fig:result_numbers}
\end{center}
\vspace{-14pt}
\end{figure*}

Task 1, on average, took 2 minutes and 48 seconds to complete (std = 2:02)(See Figure~\ref{fig:result_numbers}). This task required the participants to create a sub-user, and assign it only read permissions. Since this task was the very first interaction of the participants with the system, it would be safe to infer that the time taken to perform this task was higher than the actual time that is required to perform this task. The average completion time for Task 2 was 5 minutes and 9 seconds (std = 2:35). This time was considerably higher than the time taken to perform Task 1 (paired t test: $\mu$1 = 2:48, $\mu$2 = 5:09, p-value = 0.0001), since Task 2 required the participants to create a list ($collaborators<1,0,0>$) containing 3 email IDs, as well as a sub-user. Task 3 took an average of 5 minutes and 27 seconds to complete (std = 2:54), and was the most time consuming of all the 5 tasks. This task also needed the participants to create a list ($family<0,0,0>$, with 2 email IDs) and a sub-user, the time taken for Task 3 was only slightly higher the time taken for Task 2 (t test, t = -0.423, p-value > 0.3). This happened because a majority of participants made mistakes in this task, and had to re-perform it. We discuss the mistakes in the next subsection.

Task 4, like Task 1, also required the participants to create one sub-user, and the time taken was similar to Task 1 (t test, t = 0.613, p-value = 0.2904). In addition to Task 1, this task also required participants to add a keyword filter. Time taken to complete this task was 2 minutes and 32 seconds (std = 1:09); which is about 16 seconds lesser than the time taken to perform Task 1. This verifies our conclusion, that the first interaction of the participants with the system took more time than actually required. Further, it depicts the learning that users achieved in 4 or more interactions (depending upon the number of attempts in earlier tasks) with the system~\cite{nielsen1993usability}. Task 5 was the least time consuming, and took 1 minute and 56 seconds to complete, on average. This task required the participants to delete the sub-user created in Task 3. There was no statistically significant difference between the two groups (Gmail video first and ChaMAILeon video first) in the time taken to perform a particular task. To verify, the lowest p-value given by t test among all 5 tasks was p-value > 0.15 (p-value = 0.1603, for Task 2).

\subsubsection{Attempts to success}

The number of attempts taken to perform a task were also calculated based on sessions. Each session was counted as one attempt. All 5 tasks were successfully performed on ChaMAILeon in 1.33 attempts per task, per participant. Considering that all users were completely new to the system, and that there exist no other systems similar in functionality to ChaMAILeon, we consider this value to be very reasonable. However, during the entire process, each participant was required to log into ChaMAILeon 2.98 times per task, on average. This was because after each attempt, participants had to log in as sub-users to verify the settings they made. As discussed in the \emph{Interface feedback} section, this number could be reduced by the introduction of a ``preview" pane, which shows the participants a dynamic view of each sub-user's view of the email inbox.

\begin{table*}
    \begin{tabular}{|l|c|c|c|c|c|c|c|c|c|p{3.3cm}|}
        \hline
        {\bf Task/Permissions} & $\mathbf{R_c}$ & $\mathbf{R_{c}^{kc}}$ & $\mathbf{S_c}$ & $\mathbf{R_{nc}}$ & $\mathbf{R_{nc}^{knc}}$ & $\mathbf{S_{nc}}$ & $\mathbf{D}$ & $\mathbf{U}$ & $\mathbf{Z}$ & $\mathbf{L}$ \\ \hline
        {\bf Task} 1 & 1 & 0 & 0 & 1 & 0 & 0 & 0 & 0 & X & - \\ \hline
        {\bf Task 2} & 0 & 0 & 0 & 0 & 0 & 0 & 0 & 0 & X & collaborator<1,0,0> \\ \hline
        {\bf Task 3} & 1 & 0 & X & 1 & 0 & X & X & X & X & collaborator<1,0,0>, family<0,0,0> \\ \hline
        {\bf Task 4} & 0 & 1 (k=accounts) & 0 & 0 & 1 (k=accounts) & 0 & 0 & 0 & X & collaborator <1,0,1(k=accounts)>, family <1,0,1(k=accounts)> \\
        \hline
    \end{tabular}
\caption{Table describing success scenarios for Tasks 1 - 4. $1$ represents \emph{enabled}, $0$ represents \emph{disabled}, $X$ represents \emph{don't care condition}, and `-' represents \emph{not applicable}.}
\label{table:successscenarios}
\vspace{-7pt}
\end{table*}

Participants completed Task 1 in 1.10 attempts (std = 0.30) on an average (Figure~\ref{fig:result_numbers}). Only 3 out of the 30 participants (10\%) failed in the first attempt. Out of the 3 failed attempts, there was only one participant whose attempt failed due to incorrect settings made while configuring the sub-user for this task. This participant set the value of $R_{nc}$ = 0, while it was supposed to be 1 for success (Table~\ref{table:successscenarios}). The other two failed attempts included a participant forgetting the sub-user's password, and a participant who wanted to verify the settings twice.

Task 2, on average, took 1.50 attempts (std = 0.68) to complete successfully. This task was completed successfully on the first attempt by 18 out of the 30 participants. Common errors made during this task included $R_c$ and $R_{nc}$ being set to 1 by the participants, and participants creating the list and leaving the permissions to $collaborator<0,0,0>$ (the default value) instead of setting it to $collaborator<1,0,0>$ (Table~\ref{table:successscenarios}). A few participants also made the error of using the same passwords for sub-users created in Task 1 and Task 2, which lead to a clash, and the permissions were reset.

The most error-prone task was Task 3, where only 12 out of the 30 participants were able to successfully complete the task in the first attempt. The most common error made by the participants in this task was to miss out on setting $collaborator<1,0,0>$ (Table~\ref{table:successscenarios}). More than half of the participants were found to simply ignore the permissions for the $collaborator$ list, and only verify that the $family$ list was set to $family<0,0,0>$, which was its default value. A small percentage of participants were also found to be confused between the use of $keyword$ $filters$ and $lists$. These participants tried to set the permissions by adding the email addresses of \emph{wife} and \emph{son} in the \emph{keyword filter} field.

Task 4 turned out to be one of the easier tasks. 27 out of the 30 participants completed it successfully in the first attempt. Two out of the 3 failed attempts were caused due to the participants setting the permissions as $R_{nc}^{knc}$ = 1 (k=NULL) instead of $R_{nc}^{knc}$ = 1 (k=accounts) (Table~\ref{table:successscenarios}). We did not ~consider~ list~ permissions i.e. ~collaborator <1,0,1(k=accounts)>, and family <1,0,1(k=accounts)> for this task, since these were not a mandate for achieving success in the example we considered. The easiest of all the tasks was Task 5, which was failed by only 1 participant in the first attempt. This participant, in his first attempt, tried to remove all permissions for the sub-user created in Task 3, instead of deleting the sub-user account, which was the actual success definition for this task.

There were some common errors that the participants made across all 5 tasks, which hampered the overall number of attempts. Participants kept same passwords for different sub-users, and forgot the passwords for their sub-user accounts. These are problems which can be addressed at the system level, and can further reduce the average number of attempts for achieving success.

\subsubsection{Success Scenarios}

Upon analyzing different tasks, we found that participants made different settings to suit their needs apart from the basic settings required for the assigned tasks. Some participants preferred to remove the compose, and send / reply options to explicitly convey to the sub-user that they had restricted access. Others preferred to avoid letting their sub-users know about the restricted access, and made their system look exactly the same, with functions disabled. Participants made use of the $Z$ permissions to achieve this, and 46.6\% of the participants set this binary variable to 1, at least once during the 5 tasks. The value of this variable did not have any effect on the success or failure of any task. Different participants interpreted the settings required for Task 3 differently. This task stated that Bob needs to give access to his emails to Howard. While some participants (43.3\%) interpreted \emph{access} as both sending and reading emails, others (56.7\%) interpreted the meaning of \emph{access} as only read. Only 6.6\% of the participants set the $D$ variable to 1, and granted ``Delete" permissions to Howard in this task. Table~\ref{table:successscenarios} describes the various success scenarios as set by the participants for Task 1 through Task 4. Success of Task 5 was defined by deletion of the sub-user account created during Task 3. We use the terminology from section~\ref{sysdef} to represent the state (permissions) of the four sub-users corresponding to the first four tasks.

\begin{figure*}[ht!]
     \begin{center}
        	\subfigure[Bob's view of the mailbox]{%
            \label{fig:bob}
            \includegraphics[scale=0.28]{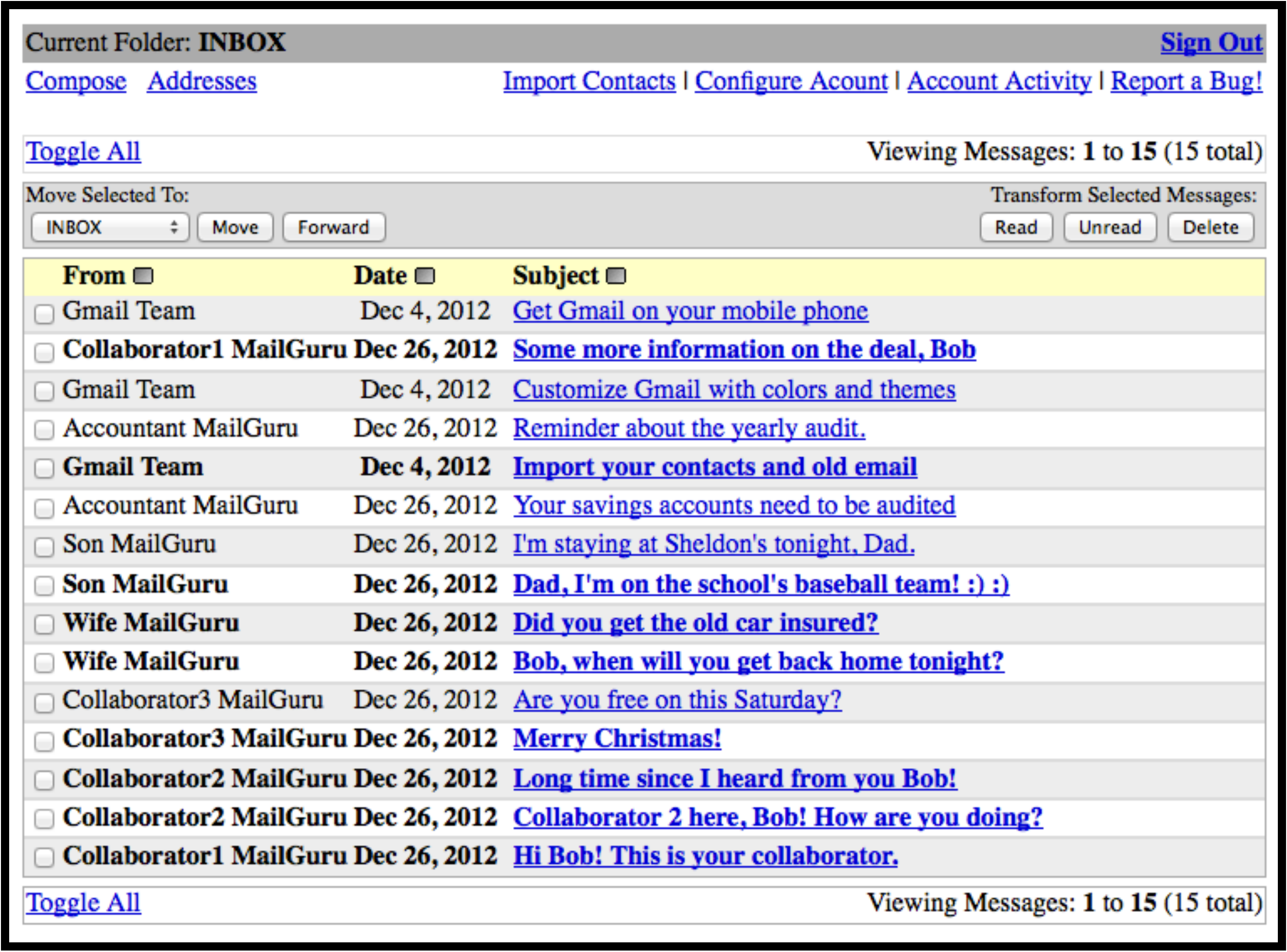}
        }%
        \subfigure[Amy's view of the mailbox]{%
           \label{fig:amy}
           \includegraphics[scale=0.28]{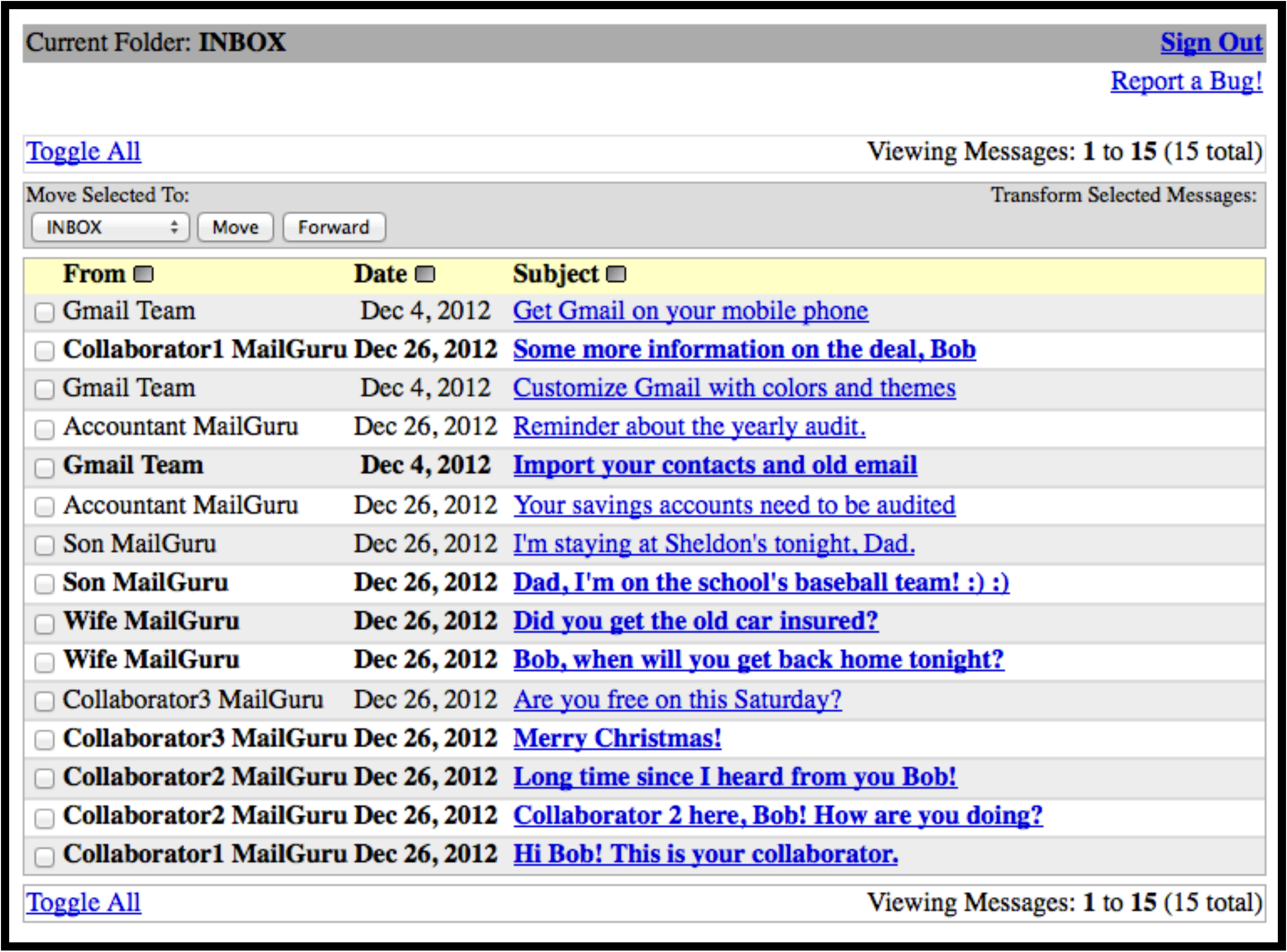}
        }\\
\subfigure[Penny's view of the mailbox]{%
           \label{fig:penny}
           \includegraphics[scale=0.27]{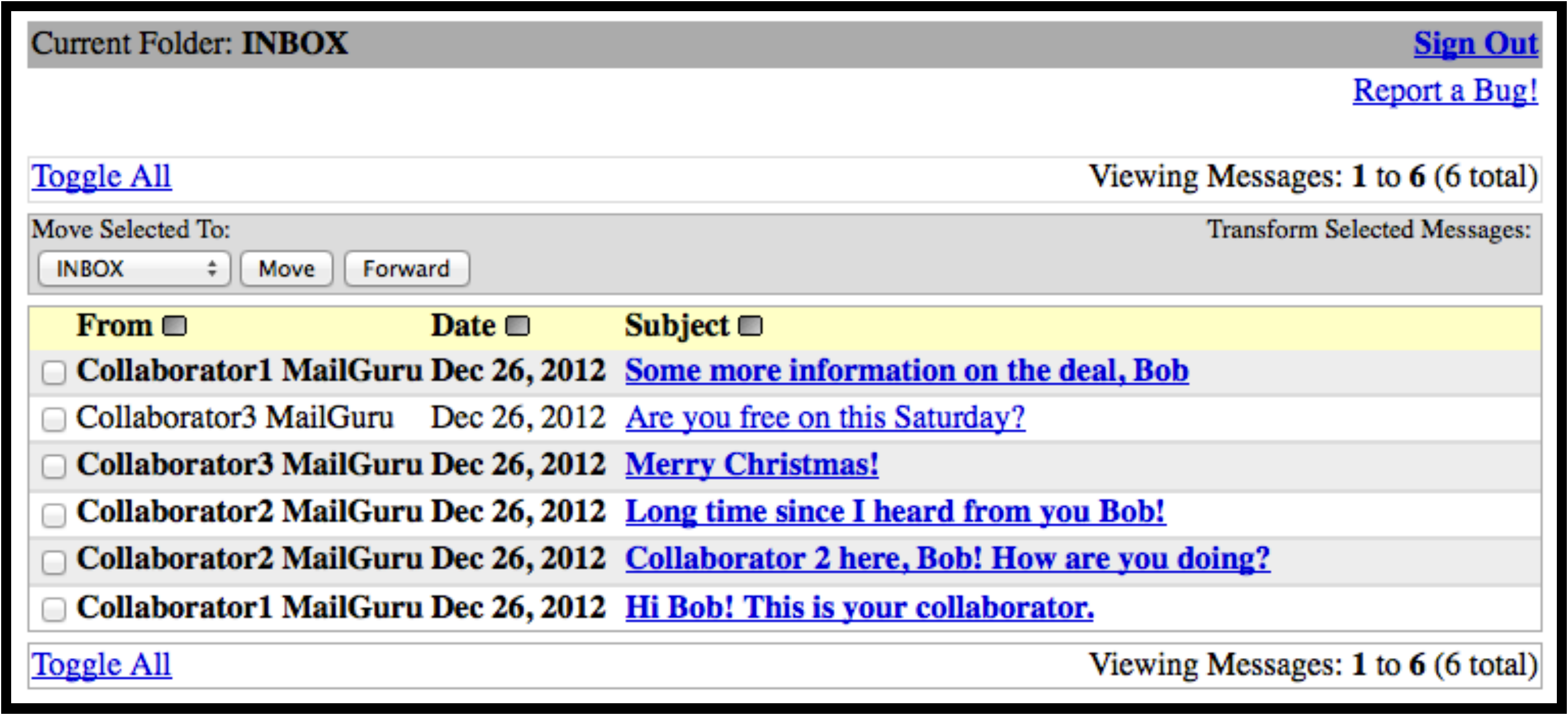}
        }
\subfigure[Howard's view of the mailbox]{%
           \label{fig:howard}
           \includegraphics[scale=0.27]{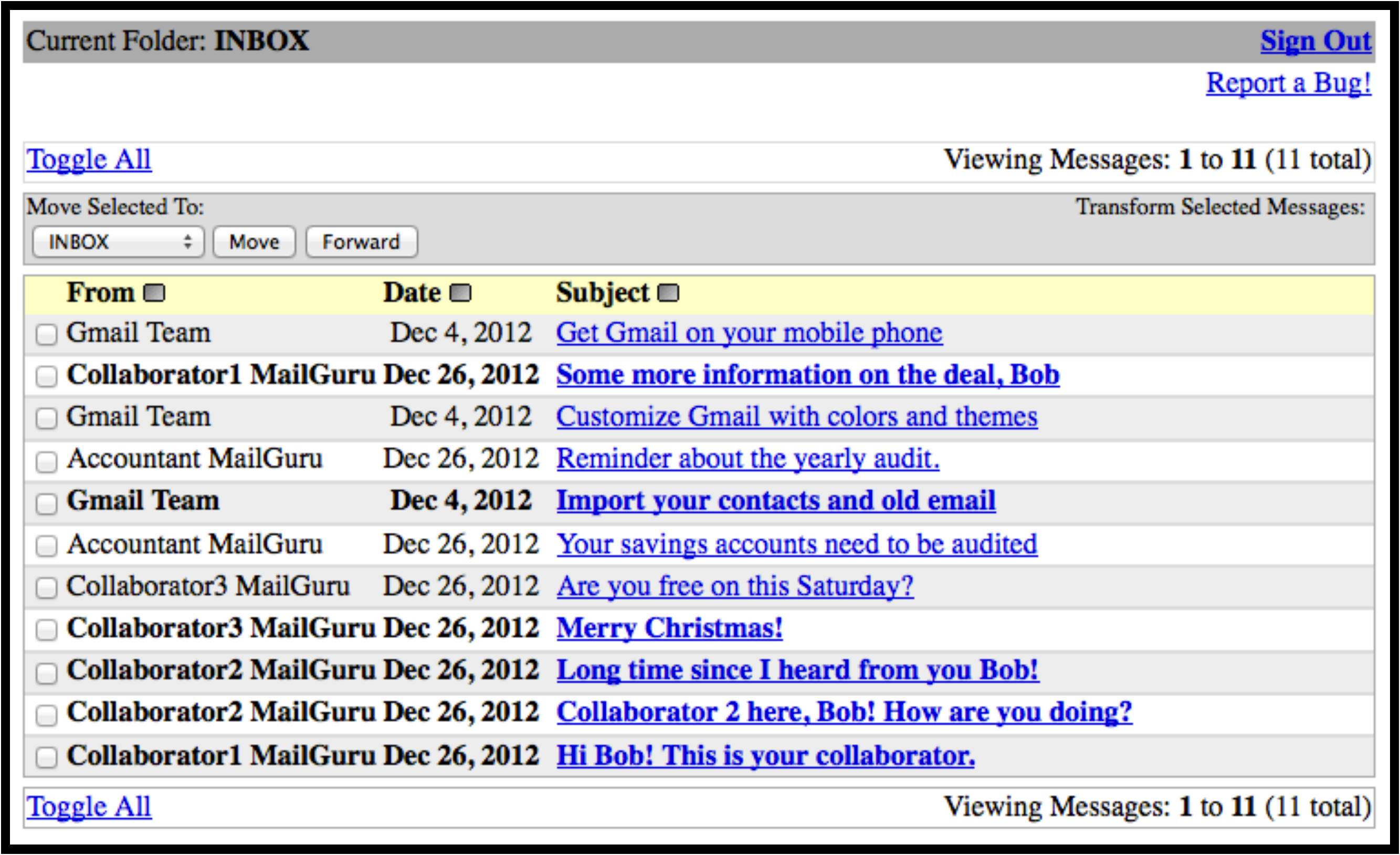}
        }
\subfigure[Stuart's view of the mailbox]{%
           \label{fig:stuart}
           \includegraphics[scale=0.29]{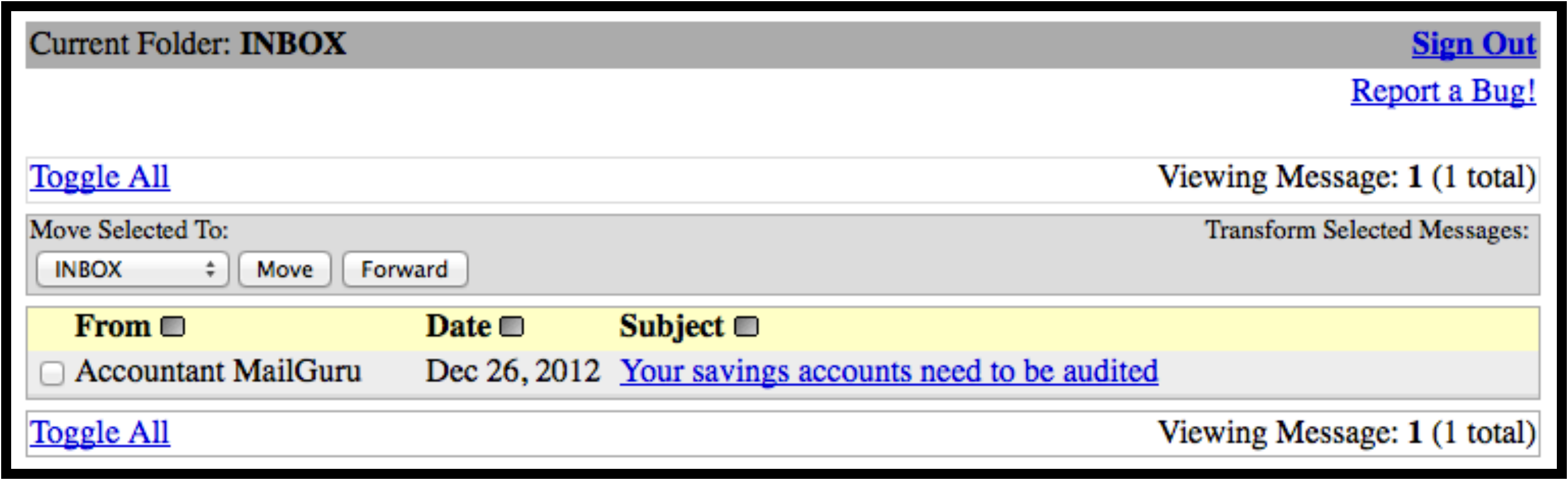}
        }
    \end{center}
\vspace{-10pt}
\vspace{-5pt}
    \caption{%
Different views of the same mailbox for different sub-users. Note that only Bob (subfigure~\ref{fig:bob}), the owner of the account gets access to the ``Configure Account", ``Account Activity", ``Import Contacts", ``Compose", and ``Addresses" features.
     }%
   \label{fig:inboxes1}
\vspace{-10pt}
\end{figure*}

In Figure~\ref{fig:inboxes1}, we present the different views of Bob's inbox, as configured for various sub-users. Bob's inbox is depicted in Figure~\ref{fig:bob}, where Bob can see all emails, and has access to all functionalities like ``Compose email", ``Configure account", ``View account activity", ``Delete email" etc. Figure~\ref{fig:amy} represents Amy's view of the inbox, corresponding to Task 1. Note that Amy has only read access to the account, and so, all other features are missing. Penny's view of the same account is represented in Figure~\ref{fig:penny}. Penny only gets to read Collaborators' emails. Figure~\ref{fig:howard} depicts Howard's view, who gets access to all except 4 emails viz. 2 from \emph{wife}, and 2 from \emph{son}. Stuart gets to see only one email, which contains the word \emph{accounts} in the subject line, as shown in Figure~\ref{fig:stuart}.
\vspace{-7pt}

\section{Discussion}

In this paper, we presented and evaluated ChaMAILeon, a system which allows users to share their emails while maintaining their privacy. Our results show that apart from certain user interface issues, ChaMAILeon is fairly useful and usable. In particular, ChaMAILeon can be useful for: (1) users whose emails need to be managed by their assistants, and (2) users who wish to maintain their privacy while sharing their passwords with their partners. ChaMAILeon can also be used in scenarios where one is forced to access her email over an insecure network, like a cyber cafe. A sub-user can be created for one's own self, and confidential emails (for example, related to health, finance, etc.) can be refrained from being transmitted over insecure, or possibly compromised networks, by getting them filtered at the server-end itself. We conducted an online survey to gather users' password sharing trends and practices, and found that a large percentage of participants (74.64\%) were uncomfortable with sharing their passwords.

To the best of our knowledge, ChaMAILeon is the first attempt to provide access control over emails. In the future, we intend to improve ChaMAILeon's user interface, fix the various bugs discovered during the controlled experiment, and conduct a more comprehensive evaluation of its usability in the real world. We also intend to study and investigate the various downsides of using ChaMAILeon. Especially under professional environments, ChaMAILeon can be exploited when a user grants access to her email account to a sub-user and asks the sub-user to impersonate her in her absence. We thus plan to conduct a comprehensive analysis of the potential pros and cons of such a system, before recommending its use in the professional environment.

Under the current settings, an implicit assumption is that the user trusts the server on which ChaMAILeon is hosted. The existing implementation of the system requires to store users' ~actual ~passwords ~(in ~encrypted form) ~on ~the ~ChaMAILeon server. Whenever a genuine sub-user logs into an account through ChaMAILeon, the system, in turn, requires to fetch emails from the Gmail server. This can only be done using the owner's actual password. We are in the process of exploring alternate ways like OAuth to avoid storing passwords. Although the concern regarding trust would not exist for organizations which run their own email servers, we envision ChaMAILeon's features to be an integral part of existing email services like Gmail, Yahoo Mail, Hotmail etc. This would completely eliminate the need for a middleware. Also, creating a large number of sub-users and passwords would, theoretically, lead to a reduction in the attack space for brute-force attacks. Associating $x$ number of passwords with one email account directly increases the chances of a successful attack by $x$ times. We would therefore suggest ChaMAILeon's users to create longer and non-trivial passwords to keep their accounts protected.
\vspace{-5pt}

\section{Guidelines}

Based on our evaluation of ChaMAILeon, and feedback from our controlled experiment, we propose some generic guidelines that could be kept in mind while designing email management systems providing role based access control mechanisms in general.


\emph{\bf Provide visualization mechanisms for verification of settings made by the users.} Users were required to log out, and log in again as sub-users themselves to verify if the settings they had made were satisfactory. This process is time consuming, and can largely hamper the chances of users adopting this feature. It is therefore advisable to give users a visual preview of the settings they have made, without them having to verify the settings manually. A quick-preview window on the sub-user settings page, containing the sub-user's view of the mailbox, can be utilized for providing this.

\emph{\bf Build intuitive interfaces where excessive information from the user is required.} It is important to create interfaces which are comforting and intuitive to users, and do not overload users with too much information~\cite{buxton2007sketching}. Participants from the controlled study were found to be overwhelmed at times, while setting permissions for sub-users due to the wide range of settings available. We advise developers to have a step-by-step interface for setting permissions, where each step contains settings for only one permission. This would reduce the amount of information presented to users at any point during the process, and would enable them to make better decisions.

\emph{\bf Include comprehensive system walk-throughs and screen-casts as learning material for the users.} Exposing users to new technology and features requires well thought out learning material to accompany the technology, in order to enable users to use it effectively. This can be achieved by building step-by-step system screen-casts, which walk users through comprehensive examples~\cite{clark2011learning}. Care should be taken to make sure that walk-through videos or animations cover the most basic and important features of the technology, and at the same time, are short enough for users to view. Users tend to skip longer and too comprehensive walk-throughs.


\emph{\bf Provide users with additional tools to assist them with newly introduced features.} Extra features (like list and sub-user features) added to email management systems introduce extra work and more things to remember (like sub-user passwords) for the users. The overhead of remembering multiple passwords becomes too much for handling email accounts securely by the users~\cite{florencio2007large}. We advise developers to provide users with software or hardware based password manager tools~\cite{al2010using,safriel2004portable}.
Further, automatic email filtering and labeling techniques~\cite{bekkerman2004automatic,provost1999naive} can be tuned and utilized to pre-generate common lists, like ``Family", ``Friends", ``Colleagues" etc. This can help reduce users' overhead further.

\vspace{-4pt}
\section{Acknowledgements}

We would like to thank all members of our research group, PreCog@IIITD. We would also like to thank Sheethal Shreedhar for helping us with the implementation of some of the earlier versions of ChaMAILeon.

%
\bibliographystyle{abbrv}
\bibliography{sigproc}  

\begin{thebibliography}{10}

\bibitem{al2010using}
H.~Al-Sinani and C.~Mitchell.
\newblock Using cardspace as a password manager.
\newblock {\em Policies and Research in Identity Management}, pages 18--30,
  2010.

\bibitem{6285845}
A.~Alarifi, H.~Tootell, and P.~Hyland.
\newblock A study of information security awareness and practices in saudi
  arabia.
\newblock In {\em Communications and Information Technology (ICCIT), 2012
  International Conference on}, pages 6 --12, june 2012.

\bibitem{bekkerman2004automatic}
R.~Bekkerman.
\newblock Automatic categorization of email into folders: Benchmark experiments
  on enron and sri corpora.
\newblock {\em Computer Science Department Faculty Publication Series}, page
  218, 2004.

\bibitem{best2010handling}
S.~Best, J.~Girouard, and T.~Snow.
\newblock Handling email communications having human delegate prepared
  summaries, Nov.~9 2010.
\newblock US Patent App. 12/942,216.

\bibitem{boyd2011social}
D.~Boyd and A.~Marwick.
\newblock Social privacy in networked publics: Teens' attitudes, practices, and
  strategies.
\newblock {\em A Decade in Internet Time: Symposium on the Dynamics of the
  Internet and Society, September 2011}, 2011.

\bibitem{bradley1958complete}
J.~Bradley.
\newblock Complete counterbalancing of immediate sequential effects in a latin
  square design.
\newblock {\em Journal of the American Statistical Association},
  53(282):525--528, 1958.

\bibitem{brooke1996sus}
J.~Brooke.
\newblock {SUS}-a quick and dirty usability scale.
\newblock {\em Usability evaluation in industry}, 189:194, 1996.

\bibitem{Butler:2012}
J.~M. Butler.
\newblock Privileged password sharing: ``root'' of all evil.
\newblock {\em SANS Whitepaper}, 2012.

\bibitem{buxton2007sketching}
W.~Buxton.
\newblock {\em Sketching user experiences: getting the design right and the
  right design}.
\newblock Morgan Kaufmann, 2007.

\bibitem{clark2011learning}
R.~Clark and R.~Mayer.
\newblock {\em E-learning and the science of instruction: Proven guidelines for
  consumers and designers of multimedia learning}.
\newblock Pfeiffer, 2011.

\bibitem{crawford2001automatic}
E.~Crawford, J.~Kay, and E.~McCreath.
\newblock Automatic induction of rules for e-mail classification.
\newblock 2001.

\bibitem{ferraiolo1995role}
D.~Ferraiolo, J.~Cugini, and D.~Kuhn.
\newblock Role-based access control ({RBAC}): Features and motivations.
\newblock In {\em Proceedings of 11th Annual Computer Security Application
  Conference}, pages 241--48. sn, 1995.

\bibitem{florencio2007large}
D.~Florencio and C.~Herley.
\newblock A large-scale study of web password habits.
\newblock In {\em Proceedings of the 16th international conference on World
  Wide Web}, pages 657--666. ACM, 2007.

\bibitem{gmail-delegation}
Google.
\newblock Set up mail delegation.
\newblock {\em
  \url{http://support.google.com/mail/bin/answer.py?hl=en&answer=138350}},
  2012.

\bibitem{Hill:2012}
K.~Hill.
\newblock Why sharing passwords with your girlfriend/boyfriend is a
  spectacularly bad idea.
\newblock {\em Forbes}, 2012.

\bibitem{hudecek2005auto}
M.~Hudecek.
\newblock Auto-forwarding and auto-delegating email folder control, Jan.~7
  2005.
\newblock US Patent App. 11/030,818.

\bibitem{hunt2008mifrenz}
T.~Hunt.
\newblock Mifrenz: Safe email for children.
\newblock {\em New Zealand Journal of Applied Computing and Information
  Technology. Volume}, 12(1), 2008.

\bibitem{hwang2012method}
J.~HWANG and G.~YANG.
\newblock Method for generating and managing a user account using a
  multi-password, Nov.~2 2012.
\newblock WO Patent 2,012,148,145.

\bibitem{Hyatt:2010}
M.~Hyatt.
\newblock Managing email with an assistant.
\newblock {\em Michael Hyatt Intentional Leadership}, 2010.

\bibitem{kumaraguru:privacy-in-india:-attitud:2012:yuqfj}
P.~Kumaraguru and N.~Sachdeva.
\newblock {Privacy in India: Attitudes and Awareness V 2.0}.
\newblock Technical report, {PreCog-TR-12-001, PreCog@IIIT-Delhi}, 2012.
\newblock http://precog.iiitd.edu.in/research/privacyindia/.

\bibitem{lenhart2011teens}
A.~Lenhart, M.~Madden, A.~Smith, K.~Purcell, K.~Zickuhr, and L.~Rainie.
\newblock Teens, kindness and cruelty on social network sites.
\newblock {\em Washington, DC, Pew Research Center}, 2011.

\bibitem{nielsen1993usability}
J.~Nielsen and J.~Hackos.
\newblock {\em Usability engineering}, volume 125184069.
\newblock Academic press San Diego, 1993.

\bibitem{notoatmodjo2009passwords}
G.~Notoatmodjo and C.~Thomborson.
\newblock Passwords and perceptions.
\newblock In {\em Proceedings of the Seventh Australasian Conference on
  Information Security-Volume 98}, pages 71--78. Australian Computer Society,
  Inc., 2009.

\bibitem{Patrick08monitoringcorporate}
A.~S. Patrick.
\newblock Monitoring corporate password sharing using social network analysis.
\newblock In {\em In International Sunbelt Social Network Conference, St. Pete
  Beach}, 2008.

\bibitem{provost1999naive}
J.~Provost.
\newblock Na{\i}ve-bayes vs. rule-learning in classification of email.
\newblock {\em University of Texas at Austin}, 1999.

\bibitem{safriel2004portable}
M.~Safriel.
\newblock Portable password manager, Mar.~26 2004.
\newblock US Patent App. 10/811,278.

\bibitem{sandhu1996role}
R.~Sandhu, E.~Coyne, H.~Feinstein, and C.~Youman.
\newblock Role-based access control models.
\newblock {\em Computer}, 29(2):38--47, 1996.

\bibitem{Singh:2007:PSI:1240624.1240759}
S.~Singh, A.~Cabraal, C.~Demosthenous, G.~Astbrink, and M.~Furlong.
\newblock Password sharing: implications for security design based on social
  practice.
\newblock In {\em Proceedings of the SIGCHI Conference on Human Factors in
  Computing Systems}, CHI '07, pages 895--904, New York, NY, USA, 2007. ACM.

\bibitem{stanton2005analysis}
J.~Stanton, K.~Stam, P.~Mastrangelo, and J.~Jolton.
\newblock Analysis of end user security behaviors.
\newblock {\em Computers \& Security}, 24(2):124--133, 2005.

\bibitem{Wilson:2002}
S.~Wilson.
\newblock Combating the lazy user: An examination of various password policies
  and guidelines.
\newblock 2002.

\bibitem{zhang2003pbdm}
X.~Zhang, S.~Oh, and R.~Sandhu.
\newblock Pbdm: a flexible delegation model in {RBAC}.
\newblock In {\em Proceedings of the eighth ACM symposium on Access control
  models and technologies}, pages 149--157. ACM, 2003.

\end{thebibliography}
%
%
\appendix

\section{Online survey questionnaire} \label{append:survey}

Except Question 2, all questions in this survey were mandatory.

\begin{enumerate}
\item Do you share your email password with one or more people?
\begin{itemize}
\item     Yes [Go to question 3]
\item     No [Go to question 2]
\end{itemize}

\item  If no, do you think such a need may arise in future?
\begin{itemize}
\item     Yes
\item     No
\end{itemize}

\item  On a scale of 1 to 5, how comfortable are you with sharing your password?
\begin{itemize}
\item     1 (very comfortable)
\item     2 (somewhat comfortable)
\item     3 (neutral)
\item     4 (somewhat uncomfortable)
\item     5 (very uncomfortable)
\end{itemize}

\item  Do you think sharing passwords may lead to unwanted consequences?
\begin{itemize}
\item     Yes. It has happened to me.
\item     Yes, but it has never happened to me.
\item     No
\item     May be
\end{itemize}
\item  How would you consider changing your password after sharing it?
\begin{itemize}
\item     I would definitely change my password immediately.
\item     I would probably change it whenever convenient.
\item     I may consider changing it some time.
\item     I would probably not change it.
\item     I would definitely not change it.
\end{itemize}

\item  Imagine, a system which allows you to control the accessibility of your emails / settings / features, when you share your password. Read the three example scenarios below, and state how you would consider using such a system.
Assume that the system is capable of providing all functionality required for the scenarios mentioned below.

Example scenario 1: You want to share your email with your spouse, but do not want him/her to see emails from a particular friend, or do not want him/her to be able to reply to any emails. How would you consider using such a system?
\begin{itemize}
\item     I would definitely use it.
\item     I would probably use it.
\item     I'm not sure.
\item     I would probably not use it.
\item     I would definitely not use it.
\end{itemize}
Example scenario 2: You are an official at a big post in a company, and receive a lot of emails every day. You wish to allow your assistant to be able to log into your account and reply to emails on your behalf. However, you want your assistant to be able to see emails sent from only specific people, and not all of them. How would you consider using such a system?
\begin{itemize}
\item     I would definitely use it.
\item     I would probably use it.
\item     I'm not sure.
\item     I would probably not use it.
\item     I would definitely not use it.
\end{itemize}

Example scenario 3: You are in a cyber cafe where you are not sure if the network is secure. You need to check your email, but you are afraid your password and email content might get compromised over the insecure network. You have an option to log into your account with a different password, such that, logging in with this password gives you very limited / restricted access to your account. How would you consider using such a system?
\begin{itemize}
\item     I would definitely use it.
\item     I would probably use it.
\item     I'm not sure.
\item     I would probably not use it.
\item     I would definitely not use it.
\end{itemize}

\item  Have you ever come across any such system, which gives you some kind of access control over emails?
\begin{itemize}
\item     Yes. I use it.
\item     Yes, I have heard of it. But I don't use it.
\item     No, I have never heard of it.
\end{itemize}

Demographic information

\item  What age group do you belong to?
\begin{itemize}
\item     13 - 18 years
\item     19 - 25 years
\item     26 - 32 years
\item     33 - 40 years
\item     41 - 50 years
\item     Above 50 years
\end{itemize}

\item  Highest educational qualification completed.
\begin{itemize}
\item     School / High School
\item     Under graduation
\item     Post graduation
\item     Ph.D.
\item     Other: 
\end{itemize}

\item  Occupation / Profession
\begin{itemize}
\item     Computer Science
\item     Engineering
\item     Medicine
\item     Science, Pharmaceuticals
\item     Law
\item     Journalism
\item     Finance
\item     Business
\item     Home maker
\item     Social sciences
\item      Other: 
\end{itemize}

\item  Gender
\begin{itemize}
\item     Male
\item     Female
\end{itemize}

\item  Which email services do you use? Tick all that you use
\begin{itemize}
\item     Google (Gmail)
\item     Yahoo! Mail
\item     Rediffmail
\item     AOL mail
\item     Hotmail
\item     Other: 
\end{itemize}

\item  Country of residence \\
If you want to take part in the lucky draw, please specify an email address or phone number where we could contact you, if you are a winner. All data collected during this survey will be anonymized and aggregated for dissemination of the results. Your answers are treated confidentially and used for research purposes only. We will not use your contact information for any other purposes but to contact you to collect your prize.

Email ID / Phone number (This is optional.)
\end{enumerate}
\section{Pre-study} \label{append:pre}

\subsection{Privacy Statement} \label{append:privacy}

{\bf Q:} What data do we collect?\\
{\bf A:} During the study we will ask you several questions related to your use of emails, and will give you a set of tasks to perform. We will take an audio-visual recording of your answers using a laptop. We will also record the screen activity during the experiment. The collected data will be anonymized and treated confidentially. This data will serve as answers to research questions and will be used solely for the purpose of this study.\\
\\
{\bf Q:} Who has access to this data?\\
{\bf A:} Access to collected data is restricted to researchers involved in the study. Your personal information will never be transferred to another party. Aggregated statistics and anonymized quotes / clips from the audio-visual recordings will be published in an academic article.\\
\\
{\bf Q:} What are my rights?\\
{\bf A:} You can interrupt or withdraw form the study at any time. You may ask to see the collected data. You have the right to demand the deletion of parts or all it.

\subsection{Questionnaire} \label{append:pre-survey}

All questions in this questionnaire were mandatory.

\begin{enumerate}

\item How many email accounts do you have?
\begin{itemize}
\item 1
\item 2 or 3
\item 4 or 5
\item More than 5
\end{itemize}
\item Which email services do you use?\\
Tick all that you use.
\begin{itemize}
\item Google (Gmail)
\item Yahoo! Mail
\item Rediffmail
\item AOL Mail
\item Hotmail
\item Other: 
\end{itemize}
\item Do you maintain separate email accounts for personal and official purposes?
\begin{itemize}
\item Yes
\item No
\end{itemize}
\item Do you share your email password with one or more people?
\begin{itemize}
\item Yes
\item No
\end{itemize}
\item How comfortable are you with sharing your password with the following:
\begin{itemize}
\item Parents
\begin{itemize}
\item Very uncomfortable
\item Somewhat uncomfortable
\item Neutral
\item Somewhat comfortable
\item Very comfortable
\end{itemize}
\item Siblings							
\begin{itemize}
\item Very uncomfortable
\item Somewhat uncomfortable
\item Neutral
\item Somewhat comfortable
\item Very comfortable
\end{itemize}
\item Spouse							
\begin{itemize}
\item Very uncomfortable
\item Somewhat uncomfortable
\item Neutral
\item Somewhat comfortable
\item Very comfortable
\end{itemize}
\item Relatives							
\begin{itemize}
\item Very uncomfortable
\item Somewhat uncomfortable
\item Neutral
\item Somewhat comfortable
\item Very comfortable
\end{itemize}
\item Friends							
\begin{itemize}
\item Very uncomfortable
\item Somewhat uncomfortable
\item Neutral
\item Somewhat comfortable
\item Very comfortable
\end{itemize}
\item Colleagues
\begin{itemize}
\item Very uncomfortable
\item Somewhat uncomfortable
\item Neutral
\item Somewhat comfortable
\item Very comfortable
\end{itemize}
\end{itemize}
\item Highest educational qualification completed.
\begin{itemize}
\item School / High School
\item Under graduation
\item Post graduation
\item Ph.D.
\end{itemize}
\item What age group do you belong to?
\begin{itemize}
\item 18 - 25 years
\item 26 - 32 years
\item 33 - 40 years
\item 41 - 50 years
\item Above 50 years
\end{itemize}
\item Occupation / Profession
\begin{itemize}
\item Computer Science
\item Engineering
\item Medicine
\item Science, Pharmaceuticals
\item Law
\item Journalism
\item Finance
\item Business
\item Home maker
\item Social sciences
\item Other: 
\end{itemize}
\item Gender *
\begin{itemize}
\item Male
\item Female
\end{itemize}
\item Have you heard of Gmail's mail delegation feature?
\begin{itemize}
\item Yes, I have used it.
\item Yes, but I have never used it.
\item No.
\end{itemize}
\end{enumerate}

\section{Controlled experiment} \label{append:control}
\subsection{Instructions}  \label{append:instruct}
\begin{enumerate}
\item Please read the tasks aloud and ask for any clarification (if you need any).
\item Think aloud. Feel free to speak out your thoughts.
\item You are required to perform these tasks on two platforms; Gmail (\url{http://gmail.com}), and ChaMAILeon (\url{http://precog.iiitd.edu.in/chaMAILeon}).
\item Please be patient with the system; it might take some time to log you in.
\item We will now play two videos; a tutorial video of Gmail's Email~ Delegation~ feature, and a tutorial video of ChaMAILeon, which will walk you through the system with some example scenarios. Please watch the videos carefully to understand how both these systems works.
\item For any help during the activity, feel free to refer back to these videos at any point of time.
\end{enumerate}

\subsection{Scenario and tasks} \label{append:post}

Imagine yourself to be Bob Smith, the CEO of Bob Construction Company. You visit your office at 9 am in the morning, and you have five tasks (plus one clean-up task) to perform in the day, related to your emails. The tasks are as follows: You may perform these tasks in any order, according to your priority.

\begin{center}
Your (Bob's) email address: bob.mailguru@gmail.com\\
Your (Bob's) password: <password>\\
\end{center} 
{\bf Task 1:}
There are some legal cases currently going on, which you are part of. You need to give access of your email to your lawyer Amy. You want Amy to be able to read all your emails, but do not want her to be able to send or delete any emails from your account.
\begin{itemize}
\item \emph{Once you have configured this, log into your account as Amy to check if you are satisfied with the settings you have made. (How different / similar is the result from your expectations?)}\\
 \end{itemize}
{\bf Task 2:}
You want your personal assistant, Penny, to keep a track of emails from a certain set of your foreign collaborators. You thus need to share your email account with her, but wish that Penny reads emails from ONLY collaborator1.mailguru@gmail.com, collaborator2.mailguru@gmail.com and collaborator3.mailguru@gmail.com
 \begin{itemize}
\item \emph{Once you have configured this, log into your account as Penny to check if you are satisfied with the settings you have made. (How different / similar is the result from your expectations?)}\\
\end{itemize}
{\bf Task 3:}
You would not be able to access your emails for a few hours during the day, and so, you want your colleague Howard to access your emails during these hours. However, you ~wish~ that Howard does not get to read emails from your wife (wife.mailguru@gmail.com) and son (son.mailguru@gmail.com)
\begin{itemize}
\item \emph{Once you have configured this, log into your account as Howard to check if you are satisfied with the settings you have made. (How different / similar is the result from your expectations?)}\\
\end{itemize}
{\bf Task 4:}
There are some financial emails that you want your accountant Stuart to handle. You thus need to give access to your email to Stuart, but you wish that he should be able to see only those emails which contain the word "accounts" in the subject line.
 \begin{itemize}
\item \emph{Once you have configured this, log into your account as Stuart to check if you are satisfied with the settings you have made. (How different / similar is the result from your expectations?)}\\
\end{itemize}
{\bf Task 5:}
You learn that all your appointments for the day have been cancelled, and you realize that you no longer need Howard to access your emails. You want to revoke Howard's access to your account.
 \begin{itemize}
\item \emph{Once you have configured this, try to log into your account as Howard to check if you are satisfied with the settings you have made. (How different / similar is the result from your expectations?)}\\
\end{itemize}
{\bf Cleanup task:} Delete all lists and sub-users you have created during the activity.\\

\noindent Email IDs for the various people to whom you (participant) will give (not give) access to:
\begin{itemize}
\item Bob's wife: wife.mailguru@gmail.com
\item Bob's Son: son.mailguru@gmail.com
 
\item Bob's Foreign Collaborators:
\begin{itemize}
\item collaborator1.mailguru@gmail.com
\item collaborator2.mailguru@gmail.com
\item collaborator3.mailguru@gmail.com
 \end{itemize}

\item Accountant: acc0untant.mailguru@gmail.com
\end{itemize}

\end{document}